\documentclass[12pt,a4paper]{article}
\usepackage[textheight=23cm,textwidth=15cm]{geometry}
\usepackage{amsmath}
\usepackage{graphicx}
\usepackage{cite}
\usepackage[american]{babel}
\usepackage{here}
\usepackage[articletitle=true]{achemso}

\begin{document}

\begin{center}

{\LARGE\bf
 Ultra-fast Spectroscopy for High-Throughput and Interactive Quantum Chemistry
}

\vspace{1cm}

{\large
Francesco Bosia\footnote{ORCID: 0000-0001-6021-7672},
Thomas Weymuth\footnote{ORCID: 0000-0001-7102-7022},
and Markus Reiher\footnote{Corresponding author; e-mail: markus.reiher@phys.chem.ethz.ch, ORCID: 0000-0002-9508-1565}
}\\[4ex]

Laboratory of Physical Chemistry, ETH Zurich, Vladimir-Prelog-Weg 2, 8093 Zurich, Switzerland

June 5, 2022

\vspace{.41cm}

\end{center}

\begin{center}
\textbf{Abstract}
\end{center}
\vspace*{-.41cm}
{\small
We present ultra-fast quantum chemical methods for the calculation of infrared and ultraviolet-visible spectra designed to provide fingerprint information during autonomous and interactive explorations of molecular structures.
Characteristic spectral signals can serve as diagnostic probes for the identification and characterization
of molecular structures. These features often do not require 
ultimate accuracy with respect to peak position and intensity, which alleviates the accuracy--time dilemma in
ultra-fast electronic structure methods. If
approximate ultra-fast algorithms are supplemented with an uncertainty quantification scheme for the detection
of potentially large prediction errors in signal position and intensity,
an offline refinement will always be possible to confirm or discard the predictions of the ultra-fast approach.
Here, we present ultra-fast electronic structure methods for such a protocol in order to obtain ground- and excited-state
electronic energies,
dipole moments, and their derivatives for real-time applications in vibrational spectroscopy and photophysics.
As part of this endeavor, we devise an information-inheritance partial Hessian approach for vibrational spectroscopy, a tailored subspace 
diagonalization approach and a determinant-selection scheme for excited-state calculations.
}

\section{Introduction}
\label{sec:intro}
The presence or absence of a diagnostic spectroscopic signal can facilitate the elucidation 
of a reaction mechanism or the design of a molecular material with specific properties and function.
For example, infrared (IR) spectroscopy and ultraviolet-visible light 
(UV/Vis) spectroscopy yield useful information about a molecular system under study.
In UV/Vis spectroscopy, the position of a peak is given by the vertical transition energy 
between different electronic states at a specific nuclear configuration. 
For vibrational spectroscopy in the harmonic approximation\cite{Wilson1955, Califano1976, Bratoz1958}, 
the position of peaks can be related to the local shape of the potential energy surface (PES). 
Intensities are then usually obtained through transition probabilities by virtue of Fermi's Golden Rule\cite{Heitler1994, Craig1998}.

The quantum chemical calculation of spectroscopic information is often more time consuming than the calculation of an electronic wave function and energy. The computational cost associated with obtaining this information will be very high if large collections or sequences of molecular structures are involved;
examples are the calculation of spectra (i) for molecular dynamics trajectories\cite{Marx2009}, 
(ii) for molecular conformer ensembles\cite{Hill2012}, and 
(iii) in high-throughput virtual screening settings. 
Furthermore, efficiency is also decisive in the framework of interactive quantum chemistry\cite{Haag2013, Haag2014, Vaucher2016, Vaucher2018} because here ultra-fast delivery of quantum chemical results is the key to interactivity.
In all these cases, a speed-up of the calculation of spectra would be very beneficial. 

The calculation of IR spectra can be accelerated by the determination of only a subset of the vibrational normal modes of a molecular system according to some criterion.
In the mode-tracking approach\cite{Reiher2003, Reiher2004, Hermann2007}, the Davidson algorithm\cite{Davidson1975} is modified to refine iteratively the normal modes that are the most similar to a set of candidate vibrations at a fraction of the cost of the 
full vibrational calculation.
A similar approach has been employed in the intensity-tracking algorithm\cite{Kiewisch2008, Kiewisch2009, Luber2009, Kovyrshin2010,Kovyrshin2012, Teodoro2018}, where the most intense vibrational transitions are selectively and iteratively optimized. In the PICVic method\cite{DosSantos2014}, normal modes are calculated with an efficient and inexpensive method and the ones deemed interesting are refined with few single-point calculations with more accurate methods. 

Molecular fragmentation was also leveraged to obtain highly accurate vibrational spectra at a fraction of the cost of a full calculation\cite{Sahu2015}.
Vibrational analysis with a partial Hessian matrix\cite{Wang2016} exploits only the block-diagonal part of the full Hessian matrix corresponding to a molecular substructure of interest that 
is evaluated and diagonalized\cite{Head1997}. This approach was successfully employed in the calculation of changes in reaction enthalpy and entropy for systems in which the changes induced by the reaction are local in nature\cite{Li2002}. The partial Hessian vibrational analysis has been extended by considering the rest of the molecular system as a collection of rigid bodies allowed to rotate and translate relative to the subsystem under scrutiny\cite{Ghysels2007}. This removed spurious negative frequencies due to the fact that the partitioned substructures were frozen in the respective relative positions. In polymer chemistry, for instance, the molecular structure is partitioned in subsystems represented by the monomer of the polymeric chain, the low-frequency vibrations are approximated 
by considering the monomers as rigid blocks, successively perturbed by the high-frequency vibrations of the monomers\cite{Durand1994, Tama2000}. In the Cartesian tensor transfer method\cite{Bour1997}, the Hessian matrix and the property tensors are efficiently calculated by fragmenting the molecular structure and assembling the resulting matrices and property tensors. Infrared and Raman spectra calculated with this tensor transfer approach are, in general, well reproduced\cite{Bieler2011}.

UV/Vis spectra are calculated by solving the linear response eigenvalue equation. Efficient methods are typically based on local approximations\cite{Kovyrshin2010}, on the reduction of the excitation space\cite{Rueger2015, Grimme2013}, and on the approximation of the required integrals\cite{Grimme2013, Bannwarth2014, Niehaus2001}. 
R\"uger and co-workers described a protocol based on a modification of time-dependent density functional theory (TD-DFT) for semi-empirical density functional tight binding (DFTB), namely TD-DFTB\cite{Rueger2015}. 
In this protocol, the excited-state linear-response eigenvalue problem is solved in a small subspace of the full excitation space. This subset is determined by an intensity criterion: determinants corresponding to single excitations from the Hartree--Fock reference determinant will be added to the subset if the dipole matrix element for this excitation exceeds a predefined threshold. 
However, the effect of this basis reduction on accuracy and reliability is difficult to foresee. 

In the simplified TD-DFT (sTD-DFT) and simplified Tamm--Dancoff Approximation (sTDA)\cite{Grimme2013, Risthaus2014, Bannwarth2014}, the calculation of the two-electron integrals in the molecular orbital basis required in
the excited-state eigenvalue problem is simplified by means of the approximation of the integrals with a multipole expansion truncated after the monopole terms. 
In this way, only partial charges and molecular orbital energy differences are needed to solve the excited-state linear-response problem\cite{Grimme2016}. This approach was also adopted for time-dependent density functional tight binding (TD-DFTB)\cite{Niehaus2001}. 
The excited-state linear-response matrix is then diagonalized in a subspace defined by all 
determinants representing single excitations in which the difference between occupied and virtual orbitals involved is lower
than the maximum energy for which the UV/Vis spectrum is calculated. Excluded basis functions that have a high off-diagonal element in the excitation matrix with basis functions included in the subset are then recovered through a perturbative approach. The accuracy of sTDA and sTD-DFT 
can be similar to that of the corresponding TDA and TD-DFT, respectively, but at a fraction of the cost\cite{Grimme2013}.

Neugebauer and co-workers developed a selective TD-DFT solver automatically removing low-lying long-range charge-transfer states\cite{Kovyrshin2012}. This allows reliably to obtain the relevant states at reduced computational cost. 
Furthermore, special hardware such as graphics processing units can accelerate excited-state calculations\cite{Isborn2011}.
Finally, methods employing a small basis (to which semi-empirical methods belong to)\cite{Grimme2016, Liu2018} offer an avenue for the accelerated calculation of both UV/Vis and IR spectra.

Approximate electronic structure methods introduce errors in the calculation of spectroscopic signals, the extent of which needs to be assessed with uncertainty quantification\cite{Sullivan2011}. We studied and developed protocols to quantify the uncertainty in the molecular properties calculated by density functionals\cite{Simm2016, Simm2017b, Proppe2017}, to propagate the effect of errors in activation (free) energy barriers from first principles to species concentrations in kinetic modeling\cite{Proppe2016, Proppe2019}, and to estimate the role of uncertainty in the parametrization of dispersion corrections\cite{Weymuth2018,Proppe2019b}. Such approaches can be extended in order to be applicable to spectroscopic signals. Although this is beyond the scope of the present work, we note that Jacob and coworkers have recently published first steps into this direction\cite{Oung2018}.

Even though these developments represent remarkable advances in the efficiency of single-spectrum calculations, none of them allows for interactive spectroscopic feedback describing structural changes of molecules in real time.
In this work, we seamlessly integrate spectroscopic calculations into the ultra-fast quantum mechanical exploration of a molecular system in an automated 
fashion. This development was driven by the desire to obtain spectroscopic information on the fly in interactive quantum chemistry\cite{Haag2013, Haag2014, Vaucher2016, Vaucher2018}. Our developments may also be beneficial for a fast analysis of molecular dynamics trajectories\cite{Marx2009}, for the calculation of spectra averaged over ensembles of molecular conformers, and in automated high-throughput calculations such as reaction network explorations\cite{Sameera2016,Dewyer2018,Simm2019,Unsleber2020}.

\section{Theory}
\label{sec:theory}

We first review the essential theory to introduce key notation. All developments presented in this section
are implemented in our open-source C++ software library for semi-empirical methods called \textsc{Sparrow}\cite{Sparrow300}. Hartree atomic units are used throughout if not otherwise stated.

\subsection{Vibrational Spectroscopy}
Vibrational peak positions are obtained as differences between energy eigenvalues of 
the time-independent nuclear Schr\"{o}dinger equation, in which the electronic energy $E_{\rm el}$  is approximated as a Taylor series expansion truncated after the second derivatives with respect to the nuclear Cartesian coordinates $\boldsymbol{R}^{\rm(c)}$. For this, the Hessian matrix $\boldsymbol{F}^{\rm(c)}$,
\begin{equation}
\label{eq:diagHessian}
 F_{ij}^{\rm(c)} = \left(\frac{\partial^2E_{\rm el}(\boldsymbol{R}^{\rm(c)})}{\partial R_i^{\rm(c)} \partial R_j^{\rm(c)}}\right) ,
\end{equation}
is calculated at a local energy minimum of the PES (the indices $i$ and $j$, respectively, refer to the atomic nuclei). 
In the basis of mass-weighted normal coordinates $\boldsymbol{R}^{\rm(q)}$ the nuclear Schr\"{o}dinger equation simplifies to\cite{Neugebauer2002}
\begin{equation}
\label{eq:nucSG}
\left( -\frac{1}{2} \nabla_{\rm nuc}^{\rm(q)\dagger}\nabla_{\rm nuc}^{\rm(q)} + \frac{1}{2}\boldsymbol{R}^{\rm(q)\dagger}\boldsymbol{F}^{\rm(q)}\boldsymbol{R}^{\rm(q)} \right) |v^{\rm tot}\rangle = E^{v^{\rm tot}}_{\rm nuc}|v^{\rm tot}\rangle ,
\end{equation}
where $\nabla_{\rm nuc}^{\rm(q)}$ is the vector corresponding to the nuclear gradient expressed in the basis of mass-weighted normal coordinates 
and $|v^{\rm tot} \rangle$ is the nuclear wave function of the system with nuclear energy $E^{v^{\rm tot}}_{\rm nuc}$ (electronic and nuclear state indices have been omitted
for the sake of simplicity). 

The Hessian matrix $\boldsymbol{F}^{\rm(q)}$ is diagonal in this representation, and the total nuclear wave function is then a product of $3N$ independent single-mode harmonic oscillator wave functions, with $N$ being the number of atomic nuclei. 
The $p$-th peak position is given by the spectroscopic wavenumber $\tilde{\nu}_p$ and 
determined by the $p$-th diagonal element $F_{pp}^{\rm(q)}$ of $\boldsymbol{F}^{\rm(q)}$,
\begin{equation}
\label{eq:diagElHessian}
 F_{pp}^{\rm(q)} = 4 \pi^2c^2\tilde{\nu}_p^2,
\end{equation}
where $c$ is the speed of light in vacuum.

Eq.~(\ref{eq:nucSG}) is only valid for a vanishing nuclear gradient. In practice, this condition is enforced by a structure optimization of the molecular system, which can require a sizeable fraction of the total computational effort. The Hessian matrix in Cartesian coordinates is then determined analytically or semi-numerically, \textit{i.e.}, as finite differences of analytical gradients. It is transformed to mass-weighted coordinates and its center of mass translation and rotational components are projected out. Subsequent diagonalization then yields the peak positions of a vibrational spectrum in this harmonic approximation according to Eq.~(\ref{eq:diagElHessian}).

In our interactive molecular exploration framework, the calculation of a vibrational spectrum is started each time the structure approaches a local minimum on the PES indicated by negligible forces on all atoms. For the detection of a local minimum, it is sufficient to have 
the quantity $G$, \textit{i.e.}, the sum of all atomic nuclear gradients, satisfy the condition
\begin{equation}
\label{eq:grad_zero}
    G = \sum_a \sqrt{ \sum_{\alpha \in \{x,y,z\}}\left(\frac{\partial  E_{\rm el}(\boldsymbol{R}^{\rm(c)})}{\partial R^{\rm(c)}_{a, \alpha}} \right)^2} \leq \epsilon_{\rm grad} ,
\end{equation}
where $\epsilon_{\rm grad}$ is the threshold below which the sum of the forces acting on the nuclei of a molecular structure is such that it is considered to be close to a local minimum. 

Note that this detection threshold $\epsilon_{\rm grad}$ can be orders of magnitude larger than the threshold usually applied
to terminate converged structure optimizations because a subsequent structure refinement will always be possible after detection
of a local minimum. In this work, $\epsilon_{\rm grad}$ was chosen to be 0.55\,hartree $\cdot$ bohr$^{-1}$, but can be modified during an exploration if deemed necessary.
Along a molecular trajectory, the calculation of a harmonic vibrational spectrum (structure optimization and frequency analysis) is initiated after the automatic detection of a local minimum. For structures that then remain close the same local minimum of a PES, no vibrational spectrum after the first one is calculated.

In interactive quantum chemical explorations significant computational savings are attainable, because
the structural distortions induced by interactive manipulations are often local in nature. 
To exploit this fact in a second approximation, we compare structures corresponding to two subsequent local minima to identify distorted molecular fragments. 
Then, only the corresponding Hessian matrix entries that are expected to change have to be updated, which will reduce the computational effort significantly.
We note that the procedure outlined in this section provides peak positions in the harmonic approximation for various types of vibrational spectroscopy such as IR, vibrational circular dichroism, Raman, and Raman Optical Activity to mention only a few.

\subsection{Infrared Intensities}
The double harmonic approximation\cite{Califano1976, Wilson1955, Bratoz1958, Neugebauer2002} is the standard approach to routinely calculate IR spectra in computational chemistry. 
Within this approximation, the generation of an IR spectrum involves two steps: the determination of peak positions as described in the previous section and the calculation of the corresponding intensities. 

The intensity of the transition associated with the wavenumber $\tilde{\nu}_p$ is given by its integral absorption coefficient $\tilde{\mathcal{A}}_{p}$.
The integral absorption coefficient is proportional to the square 
of the derivative of the molecular electric dipole moment $\boldsymbol{\mu}$ with respect to 
the $p$-th normal coordinate, $R^{\rm(q)}_p$,\cite{Neugebauer2002}
\begin{equation}
 \label{eq:abs_coef}
 \tilde{\mathcal{A}_p} = \frac{N_{\rm A}\pi}{3c^2}\left(\frac{\partial\boldsymbol{\mu}}{\partial R^{\rm(q)}_p} \right)^2 ,
\end{equation}
where $N_{\rm A}$ is Avogadro's number. In \textsc{Sparrow}, we implemented the dipole derivative with respect to the nuclear coordinates as a finite difference
for the equilibrium Cartesian coordinates $R_{k, eq}^{\rm(c)}$ according to the 3-point central difference Bickley formula,
\begin{equation}
 \left(\frac{\partial \boldsymbol{\mu}}{\partial R_k^{\rm(c)}}\right)_{R_k^{\rm(c)} = R_{k, \rm{eq}}^{\rm(c)}} \approx \frac{\boldsymbol{\mu}(R_{k, \rm{eq}}^{\rm(c)} + \Delta) - \boldsymbol{\mu}(R_{k, \rm{eq}}^{\rm(c)} - \Delta)}{2\Delta} ,
\end{equation}
where $\Delta$ is a step size chosen to be 0.01\,bohr\cite{Neugebauer2002}. 
This derivative is subsequently transformed into mass-weighted normal coordinates.
 
For single-determinant wave functions, the electric dipole moment vector is defined as the sum of the classical nuclear electric dipole moment and the expectation value of the electric dipole operator $\hat{\boldsymbol{\mu}_{\rm el}} = -\sum_b^{n} \hat{\boldsymbol{r}}_b$ for the electronic ground state. For a Slater determinant $\Phi_0$, an antisymmetrized product of $M$ molecular spin orbitals $\psi_i$,
the total molecular electric dipole moment is obtained as\cite{Szabo1996} 
\begin{align}
 \boldsymbol{\mu} &= \Big\langle \Phi_0 \Big| - \sum_{b = 1}^n \hat{\boldsymbol{r}}_b \Big| \Phi_0 \Big\rangle + \sum_a^N Z_{a}\boldsymbol{R}^{\rm(c)}_{a} \nonumber\\ 
	    &= - \sum_i^{n} \langle \psi_i | \hat{\boldsymbol{r}} | \psi_i \rangle + \sum_a^N Z_{a}\boldsymbol{R}^{\rm(c)}_{a} \nonumber\\
	    &= - \sum_\mu\sum_\nu P_{\mu\nu}\langle\chi_\mu|\hat{\boldsymbol{r}}|\chi_\nu\rangle + \sum_a^N Z_{a}\boldsymbol{R}^{\rm(c)}_{a}. \label{eq:dipole}
\end{align}
Here, the Slater--Condon rules have been exploited, the index $b$ refers to the electrons, $n$ is the total number of electrons, and $\langle\chi_\mu|\hat{\boldsymbol{r}}|\chi_\nu\rangle$ is an element of the 
dipole matrix expressed in an atomic orbital basis spanned by functions $\chi_\mu$, into which the molecular orbitals $\psi_i$
are expanded. The one-electron reduced density matrix elements $P_{\mu\nu}$ are defined in the same atomic orbital basis.
$Z_{a}$ is the nuclear charge number of the $a$-th atom.

The electronic component of the dipole can be approximated by means of a population analysis such as the Mulliken population analysis\cite{Mulliken1955}. Note that many of its known limitations\cite{Reed1985,Herrmann2005} are mitigated in a minimal basis of a semi-empirical approach. 
The electric dipole moment within DFTB can be evaluated as a sum of atomic contributions by means of a Mulliken population analysis as
\begin{equation}
\label{eq:mulliken}
    \boldsymbol{\mu} = \sum_a^N \boldsymbol{\mu}_a = \sum_a^N \boldsymbol{R}^\textrm{(c)}_a \left( Z_a - \sum_{\lambda \in a} \sum_\sigma S_{\lambda\sigma} P_{\lambda\sigma}\right) \; ,
\end{equation}
where $\lambda$ is the index for an atomic orbital basis function centered on atom $a$, the index $\sigma$ refers to any atomic orbital basis function, $\boldsymbol{S}$ is the overlap matrix with elements $S_{\lambda\sigma} = \langle \chi_\lambda | \chi_\sigma \rangle$.

\subsection{UV/Vis Spectroscopy}

\subsubsection{Linear Response Formalism}
In contrast to the solution of the nuclear Schr\"odinger equation within the harmonic approximation which refers to a local minimum region of the PES, an electronic transition can be induced at every point of a PES. An efficient but approximate method should recover qualitatively correct spectra to reliably highlight characteristic electronic structural features of the system of interest.  
The solution of the Roothaan--Hall equation in a Hartree--Fock or Kohn--Sham density functional theory (DFT) formalism yields molecular orbital coefficients as eigenvectors and the molecular orbital energies as eigenvalues. In this work, the vertical transition energy $\omega_{ia\sigma}$ from the ground state to an electronic state assumed to be characterized by a single electron substitution from the occupied orbital $i$ to the virtual orbital $a$ (both of spin $\sigma$), will be denoted by $a \leftarrow i$. It can be estimated as the difference $\Delta_{ia\sigma}$ of the orbital energy of the virtual, $\varepsilon_{a\sigma}$, and the occupied, $\varepsilon_{i\sigma}$, orbitals,
\begin{equation}
\label{eq:orb_en_diff}
    \omega_{ia\sigma} \approx \Delta_{ia\sigma} = \varepsilon_{a\sigma} - \varepsilon_{i\sigma} .
\end{equation}
This excitation energy will not be reliable in most cases. Nonetheless, it may serve as a good baseline model to improve on. 
The necessity of relaxation of the orbitals in the excited configuration required specific procedures.
The maximum overlap method\cite{Gilbert2008} relaxes the orbitals through an additional self-consistent-field calculation with the electronic occupation corresponding to the $a \leftarrow i$ excitation. Similarly, the restricted open-shell Kohn--Sham theory aims at relaxing the molecular orbitals in an excited state, but, in contrast to the previous method, does so simultaneously for a linear combination of all determinants that are spin partners in a transition, so that it provides an excited state that is a pure spin state\cite{Rohrig2003, Ziegler1977, Frank1998}.

The linear response TD-DFT (LR-TD-DFT)\cite{Runge1984, Casida1998, Burke2005} and the time-dependent Hartree--Fock (TD-HF) methods both derive from the problem of a molecular system perturbed by a small electric field. 
They lead to an eigenvalue problem\cite{Casida1995, Dreuw2005}
    \begin{equation}
    \label{eq:RPA}
     \left[ \begin{array}{cc}
             \boldsymbol{A} & \boldsymbol{B} \\
             \boldsymbol{B}^* & \boldsymbol{A}^*
            \end{array} \right]
            \left[ \begin{array}{c}
             \boldsymbol{X} \\
             \boldsymbol{Y}
            \end{array} \right]
            = \omega
            \left[ \begin{array}{cc}
             \boldsymbol{1} & \boldsymbol{0} \\
             \boldsymbol{0} & -\boldsymbol{1}
            \end{array} \right]
            \left[ \begin{array}{c}
             \boldsymbol{X} \\
             \boldsymbol{Y}
            \end{array} \right] ,
    \end{equation}
with the elements of the matrices $\boldsymbol{A}$ and $\boldsymbol{B}$ expressed as
\begin{equation}
\label{eq:a_matrix_hf}
A_{ia\sigma,jb\tau} = \delta_{ij}\delta_{ab}\delta_{\sigma\tau}\Delta_{ia\sigma} + \left(ia|jb\right) - \delta_{\sigma\tau} \left(ij|ab\right) ,
\end{equation}
and 
\begin{equation}
\label{eq:b_matrix_hf}
B_{ia\sigma,jb\tau} = \left(ia|bj\right) - \delta_{\sigma\tau} \left(ib|aj\right) ,
\end{equation}
for TD-HF, and
\begin{equation}
\label{eq:a_matrix_dft}
A_{ia\sigma,jb\tau} = \delta_{ij}\delta_{ab}\delta_{\sigma\tau}\Delta_{ia\sigma} + \left(ia|jb\right) + \delta_{\sigma\tau}\left(ia|f^{\sigma\tau}_{\rm xc}|jb\right) ,
\end{equation}
and 
\begin{equation}
\label{eq:b_matrix_dft}
B_{ia\sigma,jb\tau} = \left(ia|bj\right) + \delta_{\sigma\tau}\left(ia|f^{\sigma\tau}_{xc}|bj\right) ,
\end{equation}
for TD-DFT, where the labels $\sigma$ and $\tau$ indicate the spin part of the molecular orbitals in the excitations $a \leftarrow i$ and $b \leftarrow j$, respectively. The kernel $f^{\sigma\tau}_{\rm xc}$ represents the second derivative of the exchange--correlation functional $E_{\rm xc}$ with respect to the spin densities $\rho_\sigma$ and $\rho_\tau$, $\delta_{ia}$ is a Kronecker delta, and $\boldsymbol{X}$ and $\boldsymbol{Y}$ are the eigenvectors for the excitations and de-excitations, respectively. The two-electron integrals $\left(ia|jb\right)$ are defined as
\begin{equation}
 \left(ia|jb\right) = \iint \psi_{i\sigma}(\boldsymbol{r})\psi_{a\sigma}(\boldsymbol{r}) \frac{1}{r} \psi_{j\tau}(\boldsymbol{r}')\psi_{b\tau}(\boldsymbol{r}') \mathrm{d}^3r \mathrm{d}^3r'\mathrm{d}\sigma\mathrm{d}\tau.
\end{equation}

If real molecular orbitals are assumed, the non-Hermitian eigenvalue problem of Eq.~(\ref{eq:RPA}) can be simplified to a lower-dimensional Hermitian one\cite{Jorgensen1981, Dreuw2005},
\begin{equation}
\label{eq:RPA_Herm}
    (\boldsymbol{A} - \boldsymbol{B})^\frac{1}{2} (\boldsymbol{A} + \boldsymbol{B}) (\boldsymbol{A} - \boldsymbol{B})^\frac{1}{2} \boldsymbol{Z}=\omega^2 \boldsymbol{Z},
\end{equation}
where $\boldsymbol{Z} = (\boldsymbol{A} - \boldsymbol{B})^{-\frac{1}{2}} (\boldsymbol{X} + \boldsymbol{Y})$. If no exact exchange is 
present as in pure density functionals, the matrix $(\boldsymbol{A} - \boldsymbol{B})^\frac{1}{2}$ will be diagonal, because $\left(ia|f^{\sigma\tau}_{\rm xc}|jb\right)$ is equal to $\left(ia|f^{\sigma\tau}_{\rm xc}|bj\right)$, and its square root is easy to calculate. Where this is not the case, invoking the Tamm--Dancoff approximation\cite{Hirata1999} or working within the configuration interaction (CI) singles approximation allows for the solution of a problem of the same dimension as the one in Eq.~(\ref{eq:RPA_Herm}), but without the need to compute the expensive square root of a matrix\cite{Chantzis2013}.

The CIS and TDA are invoked by neglecting the matrix $\boldsymbol{B}$, therefore simplifying Eq.~(\ref{eq:RPA}) to
\begin{equation}
\label{eq:CIS}
    \boldsymbol{A}\boldsymbol{X}=\omega \boldsymbol{X} .
\end{equation}
 Expanding the excited states in a singly-excited-determinant or configuration state function (CSF, see below) basis causes a limitation in the description of excited states with a considerable double-excitation character in TD-DFT, TD-HF, CIS and TDA. 
For their correct description more refined models are needed such as the explicit consideration of double excitations to yield
the configuration interaction with singles and doubles excitations (CISD) wave function 
or multireference schemes\cite{Liu2018}, or by improving upon the adiabatic approximation in TD-DFT accounting for the effects of a frequency-dependent exchange--correlation kernel\cite{Maitra2004, Cave2004, Mazur2009, Mazur2011, Huix-Rotllant2011}. However, such approaches are currently out of reach 
for a high-throughput framework. Furthermore, TD-DFT based on non-hybrid exchange--correlation functionals suffers from a lack of accuracy in the description of charge-transfer states\cite{Dreuw2004, Dreuw2005}. This problem may be mitigated by range-separated functionals\cite{Risthaus2014, Dreuw2004, Henderson2008, Baer2010, Leininger1997, Iikura2001, Niehaus2012} or by identification and subsequent removal of the offending excited states\cite{Kovyrshin2010, Kovyrshin2012}.

\subsubsection{Subspace Solver}
\label{sec:subspace_solber}
In this work, we calculate the excited states with semi-empirical adaptations of TD-DFT/TDA based in a DFTB framework, \textit{i.e.}, 
TD-DFTB. In the following paragraphs, we outline the equations needed for the implementation of an iterative diagonalizer based on the Davidson algorithm\cite{Davidson1975, Liu1978} for solving Eq.~(\ref{eq:RPA_Herm}). At the end of this chapter, we summarize the equations that are specific for TD-DFTB\cite{Niehaus2001}.

In a non-orthogonal modification of the block-Davidson method\cite{Parrish2016, Furche2016}, the solution to the first few roots of the eigenvalue problems in Eq.~(\ref{eq:RPA_Herm}) and Eq.~(\ref{eq:CIS}) is approximated in an incrementally growing Krylov subspace $\Omega$ of the full space. The matrix $\boldsymbol{H}$ is $(\boldsymbol{A} - \boldsymbol{B})^\frac{1}{2} (\boldsymbol{A} + \boldsymbol{B}) (\boldsymbol{A} - \boldsymbol{B})^\frac{1}{2}$ in TD-DFTB, and $\boldsymbol{A}$ if the TDA is invoked. In contrast to the original block-Davidson method, the orthogonality of the basis functions is not enforced.
In each iteration, the product of matrix $\boldsymbol{H}$ and matrix $\boldsymbol{\Omega}$ containing the vectors $b^k$ ($k \in \{1, 2, 3, \ldots\}$) spanning the subspace $\Omega$ is calculated to obtain the so-called sigma vectors,
\begin{equation}
\label{eq:sigma}
    \boldsymbol{\sigma} = \boldsymbol{H}\boldsymbol{\Omega} .
\end{equation}
In our implementation, in the first iteration $\boldsymbol{\Omega}$ has as many rows as $\boldsymbol{H}$ and a number of columns, $C$, that is defined on input and can range from the number of desired eigen pairs to the number of columns of $\boldsymbol{H}$. The elements of the top $C$ rows of $\boldsymbol{\Omega}$ are given by 
\begin{equation}
\label{eq:init_guess}
    \Omega_{ia\sigma, jb\tau} = \delta_{ij}\delta_{ab}\delta_{\sigma\tau} + \Gamma_{ia\sigma, jb\tau} ,
\end{equation}
where $\Gamma_{ia\sigma, jb\tau}$ is a random number between $-1\cdot 10^{-2}$ and $1\cdot 10^{-2}$, and the rest of the matrix is filled by zeroes. We noticed that this choice of $\boldsymbol{\Omega}$ was not able to produce solutions characterized by an eigenvector with no overlap with the initial $\boldsymbol{\Omega}$. Therefore, we added random numbers between $-1\cdot 10^{-5}$ and $1\cdot 10^{-5}$ to the first column vector of $\boldsymbol{\Omega}$, which solved the problem. 
The matrix $\boldsymbol{H}$ is then projected onto the subspace $\Omega$ by
\begin{equation}
    \boldsymbol{\tilde{H}} = \boldsymbol{\Omega}^\dagger \boldsymbol{\sigma} ,
\end{equation}
and the subspace generalized eigenvalue problem, 
\begin{equation}
\label{eq:GEP}
  \boldsymbol{\tilde{H}}v^h = \lambda^h \boldsymbol{S} v^h ,
\end{equation}
with the overlap matrix 
\begin{equation}
  \boldsymbol{S} = \boldsymbol{\Omega}^\dagger \boldsymbol{\Omega} ,
\end{equation}
is solved, yielding the subspace eigenvector $v^h$ corresponding to the $h$-th solution of Eq.~(\ref{eq:GEP}) and the estimate for the respective eigenvalue, $\lambda^h$. In the Davidson--Liu algorithm\cite{Liu1978}, the overlap matrix is taken to be equal to the identity matrix, as the orthogonality of the vectors $b^k$ spanning the subspace $\Omega$ is enforced. In the non-orthogonal version\cite{Furche2016} this is in general not the case. In particular, the norm of the vectors $b^k$ is allowed to decrease up to the point where the overlap matrix becomes almost singular. In this case, care must be taken while solving Eq.~(\ref{eq:GEP}) as the correct solution is only guaranteed for positive-definite overlap matrices, because the first step is the Cholesky decomposition of the overlap matrix. 
Therefore, we implemented a preconditioning step to reduce the condition number of the overlap matrix and we use the simultaneous diagonalization technique\cite{Fukunaga1990} to obtain a solution to Eq.~(\ref{eq:GEP}) which is valid also for overlap matrices that are almost singular.
In this more stable implementation, Eq.~(\ref{eq:GEP}) is solved first by preconditioning the overlap matrix as proposed by Furche and co-workers\cite{Furche2016},
\begin{equation}
\boldsymbol{S'} = diag(\boldsymbol{S})^{-\frac{1}{2}} \, \boldsymbol{S} \, diag(\boldsymbol{S})^{-\frac{1}{2}} \, ,
\end{equation}
where $diag(\boldsymbol{S})^{-\frac{1}{2}}$ is the diagonal matrix containing the inverse square root of the diagonal elements of $\boldsymbol{S}$. Then, the matrix containing the eigenvectors $v^h$, $\boldsymbol{v}$, and the corresponding diagonal eigenvalue matrix $\boldsymbol{\Lambda}$ are recovered by simultaneously finding a solution to the two problems
\begin{align}
    \boldsymbol{v'}^T \, \boldsymbol{\tilde{H'}} \, \boldsymbol{v'} &= \boldsymbol{\Lambda} 
\end{align}
and
\begin{align}
    \boldsymbol{v'}^T \, \boldsymbol{S'} \, \boldsymbol{v'} &= \boldsymbol{1} \, ,
\end{align}
where
\begin{equation}
    \boldsymbol{\tilde{H'}} =  diag(\boldsymbol{S})^{-\frac{1}{2}} \, \boldsymbol{\tilde{H}} \, diag(\boldsymbol{S})^{-\frac{1}{2}} \,
\end{equation}
and
\begin{equation}
 \boldsymbol{v'} = diag(\boldsymbol{S})^{\frac{1}{2}} \boldsymbol{v}.
\end{equation}
An appropriate matrix $\boldsymbol{v}$ is found by first carrying out an eigenvalue decomposition of the matrix $\boldsymbol{S'}$ by finding the matrix $\boldsymbol{U}$ such that
\begin{equation}
  \boldsymbol{U}^T \, \boldsymbol{S'} \, \boldsymbol{U} = \boldsymbol{\Sigma}\, ,
\end{equation}
with $\boldsymbol{\Sigma}$ being a diagonal matrix whose elements correspond to the eigenvalues of $\boldsymbol{S'}$.
A transformation matrix $\boldsymbol{T'}$ is constructed,
\begin{equation}
    \boldsymbol{T '} = \boldsymbol{U}_{\rm R} \, \boldsymbol{\Sigma}^{-\frac{1}{2}}_{\rm R} \, ,
\end{equation}
where $\boldsymbol{U}_{\rm R}$ is the matrix whose columns are the columns of $\boldsymbol{U}$ corresponding to a non-zero eigenvalue and $\boldsymbol{\Sigma}^{-\frac{1}{2}}_{\rm R}$ is the diagonal matrix of the inverse square root of the non-zero eigenvalues. Notably, $\boldsymbol{U}_{\rm R}$ is a $m \times r$ matrix, $\boldsymbol{\Sigma}_{\rm R}$ a $r \times r$ matrix, with $m$ 
being the dimension of $\boldsymbol{S'}$ and $r$ its rank. This is equivalent to performing the whitening transformation in the linear space of $\boldsymbol{S}$. The matrix $\boldsymbol{\tilde{H'}}$ is transformed with $\boldsymbol{T'}$ to yield the matrix $\boldsymbol{Q}$,
\begin{equation}
    \boldsymbol{Q} = \boldsymbol{T'}^T \, \boldsymbol{\tilde{H'}} \, \boldsymbol{T'} \, ,
\end{equation}
which is in turn diagonalized to yield its eigenvector matrix $\boldsymbol{T''}$ and the diagonal matrix $\boldsymbol{\Lambda}_{\rm R}$ containing the non-zero eigenvalues of $\boldsymbol{\Lambda}$,
\begin{equation}
    \boldsymbol{T''}^T \, \boldsymbol{Q} \, \boldsymbol{T''} = \boldsymbol{\Lambda}_{\rm R}\, .
\end{equation}
The solution $\boldsymbol{v'}_{\rm R}$ is finally obtained by 
\begin{equation}
    \boldsymbol{v'}_{\rm R} = \boldsymbol{T'} \, \boldsymbol{T''} \, .
\end{equation}
This method necessitates two eigenvalue decompositions and is therefore slower than the ordinary algorithm employing a Cholesky decomposition of the overlap matrix. The main advantage, however, lies in its robustness, \textit{i.e.}, in the fact that it can handle almost singular overlap matrices. Calculations indicate that our non-orthogonal Davidson--Liu algorithm adaptation with simultaneous diagonalization is often more efficient than the ordinary Davidson--Liu algorithm.
The Ritz estimate for the eigenvector $h$ in the full space is given by
\begin{equation}
  \theta^h = \boldsymbol{\Omega} v^{h} .
\end{equation}
At this point, the residual vector $R^h$ is calculated as
\begin{equation}
    R^h = \boldsymbol{H}\theta^h - \lambda^h \theta^h = \boldsymbol{\sigma}v^h - \lambda^h \theta^h ,
\end{equation}
and a new preconditioned residual $\delta^h$, defined as
\begin{equation}
  \delta^h = \left( \overline{\boldsymbol{H}} - \boldsymbol{1}\lambda^h \right)^{-1} R^h , 
\end{equation}
is added to the subspace $\Omega$ as the new guess vector $b^{{\rm dim}(\Omega)+1}$, where $\overline{\boldsymbol{H}}$ is the matrix containing the exact or approximated diagonal of $\boldsymbol{H}$. In our implementation,
$\overline{H}_{ia\sigma,ia\sigma} = \Delta_{ia\sigma}$. The iterations are repeated until the norm of $R^h$ of the desired roots drops below a user-specified threshold. \\
In Eq.~(\ref{eq:sigma}), the matrix $\boldsymbol{H}$ needs not be stored, and only its product with each vector $b^k$ spanning $\Omega$ is needed. How this product is constructed is the main algorithmic difference between CIS, TD-DFT and TD-DFTB.

In TD-DFTB, the sigma vector is the product of the matrix $(\boldsymbol{A} - \boldsymbol{B})^\frac{1}{2} (\boldsymbol{A} + \boldsymbol{B}) (\boldsymbol{A} - \boldsymbol{B})^\frac{1}{2}$ with a trial vector $b^k$. We will provide the working equations for the method and refer 
for a detailed discussion and derivation to Refs.~\citenum{Niehaus2001, Rueger2015}. By noting that the matrix $(\boldsymbol{A} - \boldsymbol{B})$ is diagonal for the DFTB method based on DFT with a pure functional, Eq.~(\ref{eq:RPA_Herm}) becomes
\begin{equation}
\label{eq:tddftb}
    \boldsymbol{\Delta}^\frac{1}{2} (\boldsymbol{A} + \boldsymbol{B}) \boldsymbol{\Delta}^\frac{1}{2} \boldsymbol{Z} = \omega^2 \boldsymbol{Z},
\end{equation}
where $\boldsymbol{\Delta}$ is the diagonal matrix of the orbital energy differences with elements defined in Eq.~(\ref{eq:orb_en_diff}). The matrix $(\boldsymbol{A} + \boldsymbol{B})$ is given according to Eqs.~(\ref{eq:a_matrix_dft}) and (\ref{eq:b_matrix_dft}) by
\begin{equation}
\label{eq:a_plus_b}
    (\boldsymbol{A} + \boldsymbol{B})_{ia\sigma,jb\tau} =  \delta_{ij}\delta_{ab}\delta_{\sigma\tau}\Delta_{ia\sigma} + 2\left( \left(ia|jb\right)  + \left(ia|f^{\sigma\tau}_{\rm xc}|jb\right) \right) .
\end{equation}
In TD-DFTB, the integrals in Eq.~(\ref{eq:a_plus_b}) are approximated with the Mulliken approximation\cite{Niehaus2001, Niehaus2009}, and Eq.~(\ref{eq:a_plus_b}) simplifies to
\begin{equation}
\label{eq:a_plus_b_dftb}
    (\boldsymbol{A} + \boldsymbol{B})_{ia\sigma,jb\tau} =  \delta_{ij}\delta_{ab}\delta_{\sigma\tau}\Delta_{ia\sigma} + 2\left( 
    \sum_{A}\sum_B q_A^{ia\sigma} q_B^{jb\tau} \left( \gamma_{AB} + \delta_{AB} (2\delta_{\sigma\tau} - 1) m_A\right)
    \right) ,
\end{equation}
where $A$, $B$ are atom indices, $\gamma_{AB}$ is an element of the matrix $\boldsymbol{\gamma}$ containing functionals of the distance of two atoms (directly recovered from the ground-state DFTB calculation), $m_A$ is the magnetic Hubbard parameter obtained from atomic DFT calculations\cite{Niehaus2001, Ruger2016}, and the elements of the matrix of Mulliken transition charges $\boldsymbol{q}$ are defined as\cite{Niehaus2001}
\begin{equation}
    q^{ia\sigma}_{A} = \frac{1}{2} \sum_{\mu \in A} \sum_\nu \left( C^{(occ), \sigma}_{\mu i}C^{(vir), \sigma}_{\nu a} S_{\mu\nu} + C^{(occ), \sigma}_{\nu i}C^{(vir), \sigma}_{\mu a} S_{\nu\mu} \right) .
\end{equation}
In case of a closed-shell reference, the solution of Eq.~(\ref{eq:CIS}) is conveniently obtained by expressing the $(\boldsymbol{A} + \boldsymbol{B})$ matrix in the basis spanned by CSF corresponding to singlet (${}^1\Psi_{ia}$) and triplet (${}^3\Psi_{ia}$) states,
\begin{align}
\label{eq:CSF}
    {}^1\Psi_{ia} &= \frac{1}{\sqrt{2}} \left(\Phi_{ia\alpha} + \Phi_{ia\beta}\right) , \\
    {}^3\Psi_{ia} &= \frac{1}{\sqrt{2}} \left(\Phi_{ia\alpha} - \Phi_{ia\beta} \right) ,
\end{align}
where $\Phi_{ia\alpha}$ denotes a determinant obtained by the substitution of the orbital $i$ with the orbital $a$, both with spin state $\alpha$. In this representation, the $(\boldsymbol{A} + \boldsymbol{B})$ matrix is block-diagonal, and the eigenvalue problem can be split into two independent smaller problems corresponding to the singlet and the triplet excited states. 
The matrix elements in the CSF basis are derived in the supplementary information. 
If the elements are expressed in CSF basis, the spin labels $\sigma$ and $\tau$ will not be used anymore, 
because CSFs are a combination of determinants corresponding to excitations with opposite spin parts from the HF determinant, as shown in Eq.~(\ref{eq:CSF}). Hence, the sigma vectors can be efficiently calculated in matrix notation by defining the matrix $\tilde{\boldsymbol{q}} = \boldsymbol{\Delta}^\frac{1}{2} \boldsymbol{q}$ as
\begin{align}
     {}^1\boldsymbol{\sigma}^k &= \boldsymbol{b}^k \boldsymbol{\Delta}^2 + 4\tilde{\boldsymbol{q}}\boldsymbol{\gamma}\tilde{\boldsymbol{q}}^T\boldsymbol{b}^k \nonumber \\ 
  {}^3\boldsymbol{\sigma}^k &= \boldsymbol{b}^k \boldsymbol{\Delta}^2 + 4\tilde{\boldsymbol{q}}\boldsymbol{m}\tilde{\boldsymbol{q}}^T\boldsymbol{b}^k ,
\end{align}
where $\boldsymbol{m}$ is a diagonal matrix with elements $m_{AA} = m_A$. The matrix products should be carried out from right to left in order to minimize their computational cost\cite{Rueger2015}. For the solution of the TDA problem of Eq.~(\ref{eq:CIS}), the sigma vectors are given in full analogy to the full TD-DFTB problem by
\begin{align}
\label{eq:sigma_TDA}
     {}^1\boldsymbol{\sigma}_{\rm TDA}^k &= \boldsymbol{b}^k \boldsymbol{\Delta} + 2\boldsymbol{q}\boldsymbol{\gamma}\boldsymbol{q}^T\boldsymbol{b}^k \nonumber \\ 
  {}^3\boldsymbol{\sigma}_{\rm TDA}^k &= \boldsymbol{b}^k \boldsymbol{\Delta} + 2\boldsymbol{q}\boldsymbol{m}\boldsymbol{q}^T\boldsymbol{b}^k .
\end{align}

The intensity of the electronic transition $I$ is given by its oscillator strength\cite{Casida1995}
\begin{equation}
    f_I = \frac{2}{3}\omega_I \sum_{\alpha \in x,y,z} \Big| \sum_{ia\sigma} \langle\phi_{i}|\hat{r}_\alpha|\phi_a\rangle c_{ia\sigma} \Big|^2,
\end{equation}
where $\omega_I$ is the $I$-th electronic transition energy and $c_{ia\sigma} = X_{ia\sigma}$ in determinant basis for CIS and TDA, and $c_{ia\sigma} = \sqrt{\frac{\Delta_{ia\sigma}}{\omega_I}}Z_{ia\sigma}$ for the full TD-DFT or RPA problem\cite{Casida1995}. In a singlet state, the coefficients of the same spatial orbitals with opposite spin are equal, \textit{i.e.}, $c_{ia\alpha} = c_{ia\beta}$, whereas in a triplet state they are opposite, \textit{i.e.}, $c_{ia\alpha} = -c_{ia\beta}$. Consequently, the oscillator strength for triplet electronic transitions is 0. The electric dipole moment integral in the molecular orbital basis can be evaluated by approximating the integral with a Mulliken population analysis,
\begin{equation}
    \langle\phi_i|\hat{\boldsymbol{r}}|\phi_a\rangle = \sum_A \boldsymbol{R}^{(c)}_A q^{ia\sigma}_{A} .
\end{equation}

\subsubsection{Pruning the Excited-State Basis}
The matrices entering the eigenvalue problems for CIS/TDA and TD-DFTB, Eq.~(\ref{eq:CIS}) and Eq.~(\ref{eq:RPA_Herm}), are assumed to be diagonally dominant\cite{Davidson1975}. As a corollary, each basis function (\textit{i.e.}, Slater determinant or CSF) interacts considerably with only few energetically close basis functions. This fact was exploited to limit the number of basis functions into which the excited states are expanded with modest effect on the accuracy of the excitation energy, the intensity, and the character of the electronic transitions\cite{Grimme2013}.
The major contribution in the electronic transition energy for a transition dominated by the excitation $a \leftarrow i$ of spin $\sigma$ is accounted for by the orbital energy difference $\Delta_{ia\sigma}$. Therefore, one can include only the basis functions with an orbital energy difference smaller than the maximum energy the UV/Vis spectrum should capture. This strategy has the unpleasant characteristic of rapidly degrading the quality of the higher excited states, as more and more basis functions that are important for them are excluded. Grimme\cite{Grimme2013} proposed a scheme based on second-order perturbation theory to mitigate this accuracy loss: one calculates the cumulative contribution of each remaining basis function corresponding to the excitation $b \leftarrow j$ with the space of the initially included basis functions corresponding to the excitation $a \leftarrow i$. In practice, the trial basis function is included as an excited-state basis function if its cumulative contribution, 
\begin{equation}
\label{eq:pert_crit}
    E_{jb\tau}^{(2)} = \sum_{ia\sigma} \frac{| A_{ia\sigma, jb\tau} |^2}{\Delta_{jb\tau} - \Delta_{ia\sigma}} ,
\end{equation}
is larger than a certain threshold, where the matrix $\boldsymbol{A}$ is substituted with the matrix $(\boldsymbol{A} - \boldsymbol{B})^\frac{1}{2} (\boldsymbol{A} + \boldsymbol{B}) (\boldsymbol{A} - \boldsymbol{B})^\frac{1}{2}$, and the energy differences in the denominator with ${\Delta^2_{jb\tau} - \Delta^2_{ia\sigma}}$ in the full TD-DFTB problem. In the latter case, $E_{jb\tau}^{(2)}$ is expressed in units of hartree$^2$.
This technique is readily applicable in case of a TD-DFTB or TDA calculation, as the matrices $\boldsymbol{A}$ and $(\boldsymbol{A} - \boldsymbol{B})^\frac{1}{2} (\boldsymbol{A} + \boldsymbol{B}) (\boldsymbol{A} - \boldsymbol{B})^\frac{1}{2}$ can be efficiently constructed. \\

The pruning of the excited-state basis introduces an error in the vertical transition energies. We outline the derivation of this error in case of the TDA, but it is analogous for TD-DFTB. The error $\Delta E_I$ on the energy of an electronic transition $I$ is given by
\begin{equation}
    \Delta E_I = E_I^F - E_I^P ,
\end{equation}
where the basis set in which the matrix $\boldsymbol{A}$ is represented, $F$, is partitioned in two parts: (i) the set of basis functions spanning the pruned space, $P$, and (ii) the set of basis functions excluded from the pruning, $S$. 
Obtaining the transition energy in the full space, $E_I^F$, is impracticable as it would require the solution of the excited-state problem in the $F$ space, nullifying the efficiency gain from the space truncation. $E_I^F$ is approximated with second-order perturbation theory,
\begin{equation}
    E_I^F \approx E_I^{(0)} + E_I^{(1)} + E_I^{(2)} \; ,
\end{equation}
and the corrections to the energy are obtained as\cite{Sharma2017}
\begin{equation}
    E_I^{(0)} = E_I^P \;
\end{equation}
\begin{equation}
    E_I^{(1)} = \sum_{p,q \in P} c^P_{I,p} c^{P, *}_{I,q} A_{pq}^{(1)} = 0 \;
\end{equation}
and 
\begin{equation}
\label{eq:pt2_error}
    E_I^{(2)} = \sum_{s \in S} \frac{ \left(\sum_{p \in P} c^P_{I,p} A_{ps}^{(1)}\right)^2}{E^P_I - A_{ss}} \; ,
\end{equation}
where we partitioned the matrix $\boldsymbol{A}$ such that
\begin{equation}
    \boldsymbol{A} = \left[ \begin{array}{cc}
             \boldsymbol{A}^{PP} & \boldsymbol{A}^{PS} \\
             \boldsymbol{A}^{SP} & \boldsymbol{A}^{SS}
            \end{array} \right] = \left[ \begin{array}{cc}
             \boldsymbol{A}^{PP} & \boldsymbol{0} \\
             \boldsymbol{0} & \boldsymbol{0}
            \end{array} \right] + \left[ \begin{array}{cc}
             \boldsymbol{0} & \boldsymbol{A}^{PS} \\
             \boldsymbol{A}^{SP} & \boldsymbol{A}^{SS}
            \end{array} \right] = \boldsymbol{A}^{(0)} + \boldsymbol{A}^{(1)}\; ,
\end{equation}
and $c_{I,i}^P$ is the coefficient with which the $i$-th basis function enters in the electronic transition $I$ calculated in the pruned space $P$. An estimate of the error introduced by the pruning is therefore given by
\begin{equation}
    \Delta E_I = E_I^F - E_I^P \approx E_I^{(2)} \;.
\end{equation}
After the solution of the excited-state problem in the pruned space $P$, obtaining a measure for the error is therefore convenient, as the evaluation of the matrix elements needed in Eq.~(\ref{eq:pt2_error}) is efficiently carried out analogously to Eq.~(\ref{eq:sigma_TDA}).

In direct methods where a full four-index transformation of the integrals from an atomic orbital basis to a molecular orbital basis is too expensive, as in the case of CIS or TD-DFT, one must develop a contraction scheme that allows to benefit from the excited-state space pruning. One of the computational bottlenecks within an iteration of the Davidson algorithm is the contraction of the two-electron integrals with the pseudo-density matrix in the atomic orbital basis for the generation of the sigma vectors. Since the atomic orbital basis is unaffected from the pruning described above, we employ a partial transformation of the basis in which the two-electron integrals are expressed. The benefit is twofold: first, the number of integrals is decreased from $\mathcal{O}((O+V)^4)$ to $\mathcal{O}((O+V)^2(OV))$, where $O$ is the number of occupied orbitals and $V$ is the number of virtual orbitals. Second, it allows for the pruning of the indices expressed in the molecular orbital basis. Both factors accelerate the contraction of the two-electron integrals with the trial vectors. This approach consists in the transformation of two of the four indices of the Coulomb $(\mu\nu|\lambda \sigma)$ and exchange $(\mu \sigma|\lambda\nu)$ integrals from the atomic orbital basis into a basis formed by pairs of molecular orbitals corresponding to an electronic transition $a' \leftarrow i'$ of spin $\tau$ still present after pruning,
\begin{align}
\label{eq:1etrans}
    (\mu\nu|i'a') &= \sum_{\lambda\sigma} C^{(occ), \tau}_{i'\lambda} C^{(vir), \tau}_{a'\sigma} (\mu \nu | \lambda \sigma) \nonumber\\
    (\mu a'|i'\nu) &= \sum_{\lambda\sigma} C^{(occ), \tau}_{i'\lambda} C^{(vir), \tau}_{a'\sigma} (\mu \sigma | \lambda \nu) .
\end{align}

\section{Computational Methodology}
\label{sec:comp_method}
In this work, DFTB3\cite{Gaus2011} (with the parameter set ``3ob-3-1'') was employed for ground-state calculations and the evaluation of the Hessian matrices for IR spectroscopy. In the TD-DFTB method, no specific DFTB3 term is included, and the excited-state calculation is limited to a second order expansion with respect to the density and, therefore, to terms specific to the DFTB2\cite{Elstner1998} method. However, this was shown not to affect the accuracy\cite{Nishimoto2015}.

Double-harmonic IR spectra were calculated in local minima of the PES. Along a trajectory, the exact local minimum was seldom reached. Therefore, we started an IR spectrum calculation as the molecular structure got close to a minimum, \textit{i.e.}, the sum of the atomic forces was smaller than a threshold $\epsilon_{\rm grad} = 0.55$\,hartree$\cdot$bohr$^{-1}$. At this point, the structure was optimized, where not otherwise specified, with the ``Very Tight'' convergence criteria described in Table~\ref{tab:conv_crit} and a frequency analysis was carried out\cite{Bosia2020}. The elements of the Hessian matrix were obtained by a seminumerical procedure with a step size of 0.01\,bohr.
For the partial Hessian approach, we devised an iterative algorithm that would avoid fitting to parts of the molecule that have been distorted. The algorithm fits the molecular structure corresponding to the current local minimum to the one of the preceding local minimum iteratively. During each iteration, the nuclei are classified depending on the RMSD given by a quaternion fitting procedure\cite{Coutsias2004} in three sets: one with the nuclei whose coordinates have abundantly diverged between the two structures (in this work, this is defined as nuclei with a RMSD determined by the fit exceeding 1.0\,bohr), one with nuclei that have an RMSD smaller than $\epsilon_{\rm RMSD}$, and the rest. The nuclei that have abundantly diverged are removed from the fitting set, and the next iteration is started. This procedure is repeated until the set of the nuclei with a RMSD smaller than $\epsilon_{\rm RMSD}$ does not change anymore.

How often the excited states are calculated along the trajectory is decided on by the user at the start of an exploration. In UV/Vis spectroscopy, we exploit algorithmic acceleration of the excited-states linear-response problem through a non-orthogonal implementation of the Davidson--Liu algorithm. 

We studied our approximations at the example of three exemplary trajectories. The trajectories are available in the supplementary information in the concatenated XYZ format. The trajectories \textbf{T1} and \textbf{T2} involve long-chained enols undergoing an interactively induced keto--enol tautomerism. In \textbf{T1}, the reaction is induced at one end of the aliphatic chain. In \textbf{T2}, the reaction is induced in the middle of the aliphatic chain. During the interactive exploration session, an external force was applied by the user on the oxygen-bound hydrogen atom of the enol, in order to break the bond with the oxygen and build one with the carbon. The electronic structure in the interactive exploration was calculated with the PM6 method\cite{Stewart2007}, and both trajectories were refined with the DFTB3 method in a B-Spline optimization\cite{Vaucher2018}.

\begin{figure}[hbpt]
    \centering
    \includegraphics[width=0.5\textwidth]{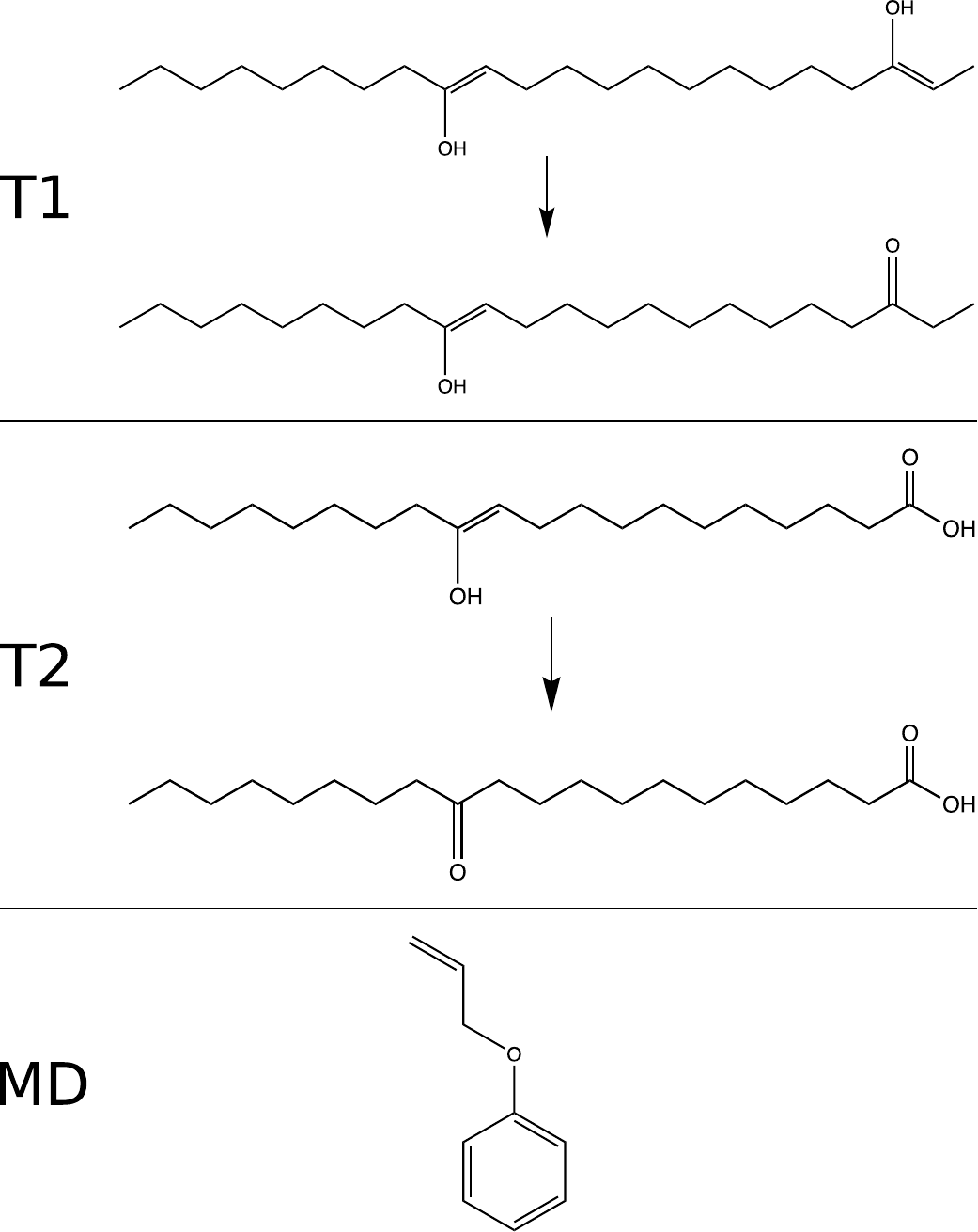}
    \caption{Lewis structures involved in the three trajectories labeled as \textbf{T1}, \textbf{T2}, and \textbf{MD} in this work. }
    \label{fig:trajectories}
\end{figure}

The trajectory \textbf{MD} was generated by a force-field molecular dynamics simulation of allylphenylether \textit{in vacuo} with the leap-frog algorithm and an integration step of 1\,fs. One structure was recorded every 250 steps. The force-field parameters were optimized in a system-focused fashion according to the \textsc{SFAM} method\cite{Brunken2020} with RI/PBE-D3BJ/def2-SVP\cite{Perdew1996, Perdew1997, Weigend2005, Grimme2010} and the def2/J auxiliary basis\cite{Weigend2006} and are available in the supplementary information. The molecular dynamics simulation was carried out with the \textsc{SCINE} software package\cite{Bosia2020}. Initial velocities were sampled from a Maxwell--Boltzmann distribution at 300\,K and the system was coupled to a Berendsen thermostat\cite{Berendsen1984} at 300\,K with a coupling time of 10\,fs. The Berendsen thermostat is known not to create a canonical ensemble and to suffer from the ``flying ice cube'' effect\cite{Harvey1998}. However, both limitations are not relevant for the scope of this work, as no thermodynamic data are extracted from the molecular dynamics simulation.

With $\epsilon_{\rm grad} = 0.55$\,hartree$\cdot$bohr$^{-1}$, local minima are detected in the trajectory \textbf{T1} at the first and 177-th structures, and for the trajectory \textbf{T2} at the first and 184-th structure. These structures are labelled as \textbf{T1.I}, \textbf{T1.II}, \textbf{T2.I}, \textbf{T2.II}, respectively. The acronym of the optimization tightness is added as a suffix to indicate the convergence criterion of the structure optimization. For the convergence criteria summarized in Table~\ref{tab:conv_crit}, the optimized structures for the first minimum of the trajectory \textbf{T1} are labelled as \textbf{T1.I.N}, \textbf{T1.I.VL}, \textbf{T1.I.L}, \textbf{T1.I.M}, \textbf{T1.I.T}, and \textbf{T1.I.VT} in order of increasing tightness of the structure optimization.

In order to assess the reliability of our approach for the calculation of UV/Vis spectra, we compared UV/Vis spectra of 200 structures evenly spaced along the \textbf{MD} trajectory, \textit{i.e.}, every twentieth structure in the trajectory, calculated with the TD-DFTB method with the ones calculated with the linear-response SCS-CC2 method\cite{Christiansen1995, Hellweg2008} with default spin component scaling constants and the cc-pVTZ\cite{Dunning1989, Woon1993} as implemented in Turbomole 7.4.1\cite{Ahlrichs1989}. A comparison with the more accurate, but prohibitively costly equation-of-motion CC3 method implemented in the $e^T$ 1.0.7 program\cite{Folkestad2020}, demonstrated the accuracy of the SCS-CC2 method as a reference for the description of valence excitations in the system at hand (data available in the supplementary information). The UV/Vis spectra were generated by convolution of the stick-spectrum obtained on the linear-response calculation with a Lorentzian with full width at half maximum of 0.3\,eV. For TD-DFTB, the first 30 excited states were calculated, for linear-response SCS-CC2, the first 10 excited states were calculated. The difference in the number of calculated excited states can be explained with ``ghost'' states: misrepresented low-lying charge transfer states, present in GGA exchange-correlation density functionals upon which the TD-DFTB formalism is based\cite{Kovyrshin2012, Goerigk2010, Sundholm2003, Neugebauer2005}. Long-range-corrected TD-DFTB\cite{Niehaus2012, Bold2020} could mitigate this condition.

We compared the efficiency of our implementation against the one of the DFTB+ 18.2 software package\cite{Hourahine2020} for the calculation of the first 30 excited states with an initial guess space of 30 vectors of the first structure of the \textbf{T1} trajectory. A reference calculation with the DFTB2 model was carried out, and the excited states were calculated with the TD-DFTB2 model with both DFTB+ and \textsc{Sparrow}. The DFTB+ input file as well as the input for the calculation with \textsc{Sparrow} are available in the supplementary information in a compressed folder.

Normal modes were matched with a linear sum assignment\cite{Kuhn1955} as implemented in the SciPy 1.4.1 Python package\cite{Scipy2020} with the element-wise absolute value of the Duschinsky matrix $|\boldsymbol{R}^{(q)\dagger}\boldsymbol{R}^{(q)}|$ as score matrix, with exception of the normal modes calculated in section~\ref{sec:approx_struct_opt}, which were matched according to their energetic ordering, as the different molecular structures involved made the previous assignment unreliable. 

All calculations were performed on a computer equipped with an Intel Xeon E-2176G CPU (3.70\,GHz base frequency) on 6 parallel threads. A very limited amount of virtual memory is needed for the calculation of systems  of this size with semi-empirical methods.

\section{Results}

In this section we analyze the reliability of the approximations needed to carry out ultra-fast calculations of IR and UV/Vis spectra.

\subsection{Infrared Spectroscopy}

First, we inspect the reliability of two prototypical semi-empirical models, PM6 and DFTB3, for the calculation of IR spectra, by evaluating the absolute deviation of the Hessian matrix elements of \textbf{T2.I} calculated with DFT (PBE0/def2-TZVP/D3\cite{Adamo1999,Weigend2005, Grimme2010}, implemented in Orca 4.2.0\cite{Neese2012,Neese2018}), DFTB3 and PM6. Furthermore, the vibrational frequencies of \textbf{T2.I} obtained with these methods were compared to each other. These calculations were carried out on the structure optimized with DFT. Both the Orca input file and the coordinates of the structure analyzed are available in the supplementary information. It is obvious from Fig.~\ref{fig:freq_comparison}, especially when comparing the high-wavenumber modes, that DFTB3 is a good candidate semi-empirical method for the calculation of infrared spectra of quality similar to the ones calculated with DFT for the organic molecule under study. The superiority of DFTB3 over the PM6 model is corroborated by the analysis of the absolute deviations of the Hessian matrix elements between the different methods shown in the supplementary information, where the difference between the Hessian matrices calculated with PM6 and either DFT or DFTB3 is considerably larger than the one between the Hessian matrices calculated with DFT and DFTB3. Hence, DFTB3 was chosen for all IR spectroscopy calculations in this work. 

\begin{figure}[hbpt]
    \centering
    \includegraphics[width=0.9\textwidth]{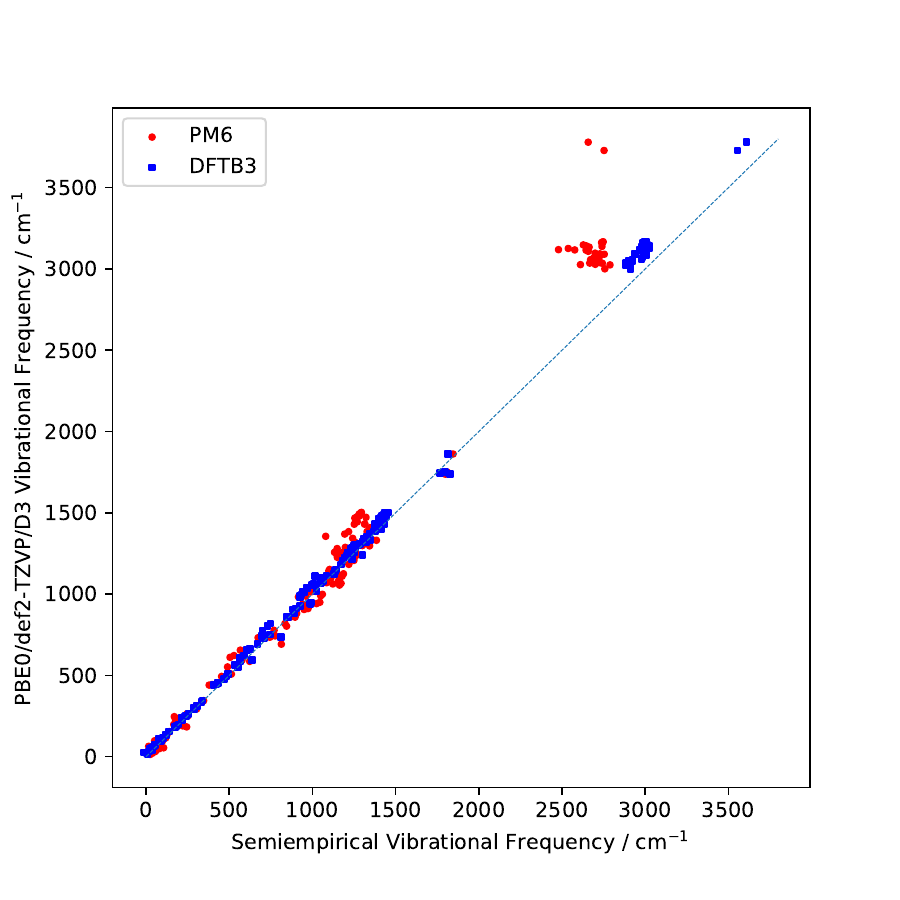}
    \caption{Comparison of the vibrational frequencies calculated for the  \textbf{T2.I} structure optimized with DFT (PBE0/def2-TZVP/D3), PM6 and DFTB3. Red circles refer to the comparison of DFT and PM6 vibrational frequencies, blue boxes to the one of DFT and DFTB3. The diagonal striped line indicates the identity. The normal modes were matched with a linear sum assignment with the absolute value of the Duschinsky matrix as score matrix.}
    \label{fig:freq_comparison}
\end{figure}

\begin{figure}[hbpt]
    \centering
    \includegraphics[width=0.9\textwidth]{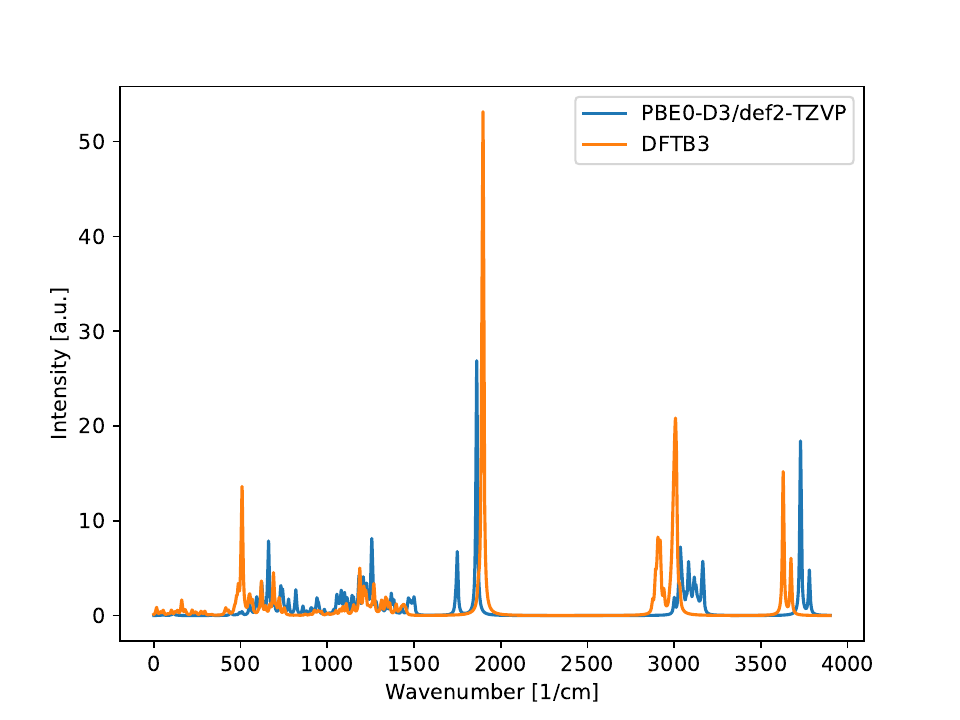}
    \caption{Comparison of the IR spectra of \textbf{T2.I}, calculated with DFT (PBE0/def2-TZVP/D3) and DFTB3.}
    \label{fig:spec_comparison}
\end{figure}

Besides the vibrational frequencies, it is also important that the IR intensities are described sufficiently well. To investigate this, we compare the vibrational spectrum of \textbf{T2.I} as calculated with DFT (PBE0/def2-TZVP/D3) to the same spectrum obtained from DFTB3 (see Fig.~\ref{fig:spec_comparison}). As we can see, DFTB3 is not always capable to quantitavely recover the DFT results. For example, the intensities of the normal mode slightly below 2000\,cm$^{-1}$ and the peak at about 3000\,cm$^{-1}$ are overestimated by a factor of two and more. Despite these obvious shortcomings, however, DFTB3 is able to qualitatively recover the entire vibrational spectrum.

Next, we study the two approximations which affect the IR spectrum calculations. First, we assess the loss in accuracy of the position of the peaks and in the elements of the Hessian matrix if the molecular structure is optimized with loose convergence criteria. Second, we evaluate the partial Hessian approach. 

\begin{table}[hbpt]
    \centering
        \caption{Mean computational time and number of DFTB3 steps required for structure optimization over all traversed minima on the PES without any of the approximations presented in this work. Times are separated in the one required to calculate the structure optimizations and the one required to calculate and diagonalize the Hessian matrices and to evaluate the dipole gradient, the sum of which is under the column named ``Time for Hessian matrix''. Timings are presented for the \textbf{T1} and \textbf{T2} trajectories and are given as mean of the calculation time of 3 calculations $\pm$ standard deviation over all minima in the trajectory. Significant digits are given by the standard deviation: if it is larger than 2.5 multiplied by the appropriate power of ten, then it is rounded to the first digit, otherwise to the second one. The convergence criteria are listed in Table~\ref{tab:conv_crit} under the ``Tight'' optimization profile. In the structure optimization in the first minimum of the trajectory \textbf{T2} the internal coordinates break down after iteration 19. Afterwards, the optimization is resumed in Cartesian coordinates.}
    \vspace{0.1cm}
    \begin{tabular}{l|ccc}
    \hline\hline
                &        Time for             &  Number of        & Time for    \\
         System & structure optimization [ms] &      steps        & Hessian matrix [ms] \\
         \hline
         \textbf{T1} &                 &     &                   \\
          I. minimum &  $4830 \pm 30$  & 111 &   $2540 \pm 140$  \\
         II. minimum &  $4580 \pm 100$ & 113 &   $2341 \pm 7$    \\
         \hline
         \textbf{T2} &                 &     &                   \\
          I. minimum &  $18500 \pm 120$  & 805 &   $1880.7\pm 1.5$ \\
         II. minimum &  $2886 \pm 23$  & 79  &   $2200 \pm 500$  \\
         \hline\hline
    \end{tabular}
    \label{tab:time_normal}
\end{table}

The trajectories \textbf{T1} and \textbf{T2} represent challenging targets for ultra-fast infrared spectroscopy because of their size. For these calculations, $\epsilon_{\rm grad} = 0.55$\,hartree$\cdot$bohr$^{-1}$ results in two minima being detected along the trajectories corresponding to the start and end states of the system as shown in Fig.~\ref{fig:trajectories} for \textbf{T1} and \textbf{T2}. The energies and gradients along these trajectories are available in the supplementary information.

\subsubsection{Approximate Structure Optimization}
\label{sec:approx_struct_opt}

The structure optimization represents a sizeable fraction of the computational effort to obtain an IR spectrum, as shown in Table~\ref{tab:time_normal}. Therefore, we explored to what extent a partial structure optimization affects the elements of the Hessian matrices of the structures with different structure optimization convergence criteria and the position of the peaks in the respective IR spectra. To this aim, we define five optimization profiles, \textit{i.e.}, different sets of criteria governing the convergence of the structure optimization, summarized in Table~\ref{tab:conv_crit}. 
\begin{table}[hbpt]
    \centering
    \caption{Convergence criteria in atomic units for the different optimization profiles, defined in the main text. The profile ``None'' indicates no optimization. ``Max Step'' and ``RMS Step'' are the maximum deviations of any Cartesian coordinate and the root mean square deviation of the Cartesian coordinates vector between two iterations. ``Gradient'' is the nuclear gradient of the total energy. ``Max'' and ``RMS'' have the same meaning as above. $\Delta$ is the variation of the total energy between two iterations. \# indicates the number of criteria to satisfy besides $\Delta$ in order to reach convergence. All values are given in atomic units.}
    \vspace{0.1cm}
    \begin{tabular}{l|cccccc}
        \hline \hline
        Optimization  & Max & RMS & Max & RMS & $\Delta$ & \# \\
        profile&Step&Step&Gradient&Gradient& &\\
        \hline
         None & - & - & - & - & - & - \\
         Very Loose & $1 \cdot 10^{-2}$ & $5 \cdot 10^{-2}$ & $5 \cdot 10^{-3}$ & $1 \cdot 10^{-2}$ & $1 \cdot 10^{-4}$ & 2 \\
         Loose & $5 \cdot 10^{-3}$ & $1 \cdot 10^{-2}$ & $1 \cdot 10^{-3}$ & $5 \cdot 10^{-3}$ & $1 \cdot 10^{-5}$ & 2\\
         Medium & $1 \cdot 10^{-4}$ & $5 \cdot 10^{-3}$ & $5 \cdot 10^{-4}$ & $1 \cdot 10^{-4}$ & $1 \cdot 10^{-6}$  &2\\
         Tight & $1 \cdot 10^{-4}$ & $5 \cdot 10^{-4}$ & $5 \cdot 10^{-5}$ & $1 \cdot 10^{-5}$ & $1 \cdot 10^{-7}$ & 3\\
         Very Tight & $2 \cdot 10^{-5}$ & $1 \cdot 10^{-5}$ & $2 \cdot 10^{-5}$ & $1 \cdot 10^{-5}$ & $1 \cdot 10^{-7}$ & 4\\
         \hline \hline
    \end{tabular}
    \label{tab:conv_crit}
\end{table}

\begin{table}[hbpt]
    \centering
        \caption{Computational times required for the DFTB3 structure optimization of \textbf{T1.II} and \textbf{T2.II} with different convergence criteria. Timings are given as mean of the calculation time of 3 calculations $\pm$ standard deviation. Significant digits are given by the standard deviation: if it is larger than 2.5 multiplied by the appropriate power of ten, then it is rounded to the first digit, otherwise to the second one.}
%    \small
%    \tabcolsep=0.10cm
    \vspace{0.1cm}
    \begin{tabular}{l|ll}
            \hline \hline
            &  Optimization  & Optimization   \\
         Structure & profile & time [ms]  \\
               \hline
         \textbf{T1.II} & None       & -                \\
                     & Very Loose & $535.7 \pm 2.0$  \\
                     & Loose      & $1738 \pm 6$     \\
                     & Medium     & $3256 \pm 7$     \\
                     & Tight      & $4580 \pm 100$   \\
                     & Very Tight & $5473.3 \pm 2.5$ \\
                     \hline
        \textbf{T2.II}  & None       & -                \\
                     & Very Loose & $503.7 \pm 2.3$  \\
                     & Loose      & $1524.0 \pm 2.2$ \\
                     & Medium     & $2357 \pm 7$     \\
                     & Tight      & $2856 \pm 21$    \\
                     & Very Tight & $4710 \pm 160$   \\
        \hline \hline

    \end{tabular}
    \label{tab:approx_opt_timings}
\end{table}

We assessed the viability of carrying out an approximate structure optimization before calculating an IR spectrum by two criteria: the gain in efficiency and the loss in accuracy with respect to performing a full structure optimization. We summarized the mean times needed to carry out a structure optimization for \textbf{T1.II} and \textbf{T2.II} in Table~\ref{tab:approx_opt_timings}. \textbf{T1.I} and \textbf{T2.I} are detected at the first structure along the trajectories, and their comparison is therefore skewed by the different starting situation of the trajectories. The acceleration factors range from 1.2 to more than a magnitude. 
\begin{figure}[hbpt]
    \centering
    \includegraphics[width=\textwidth]{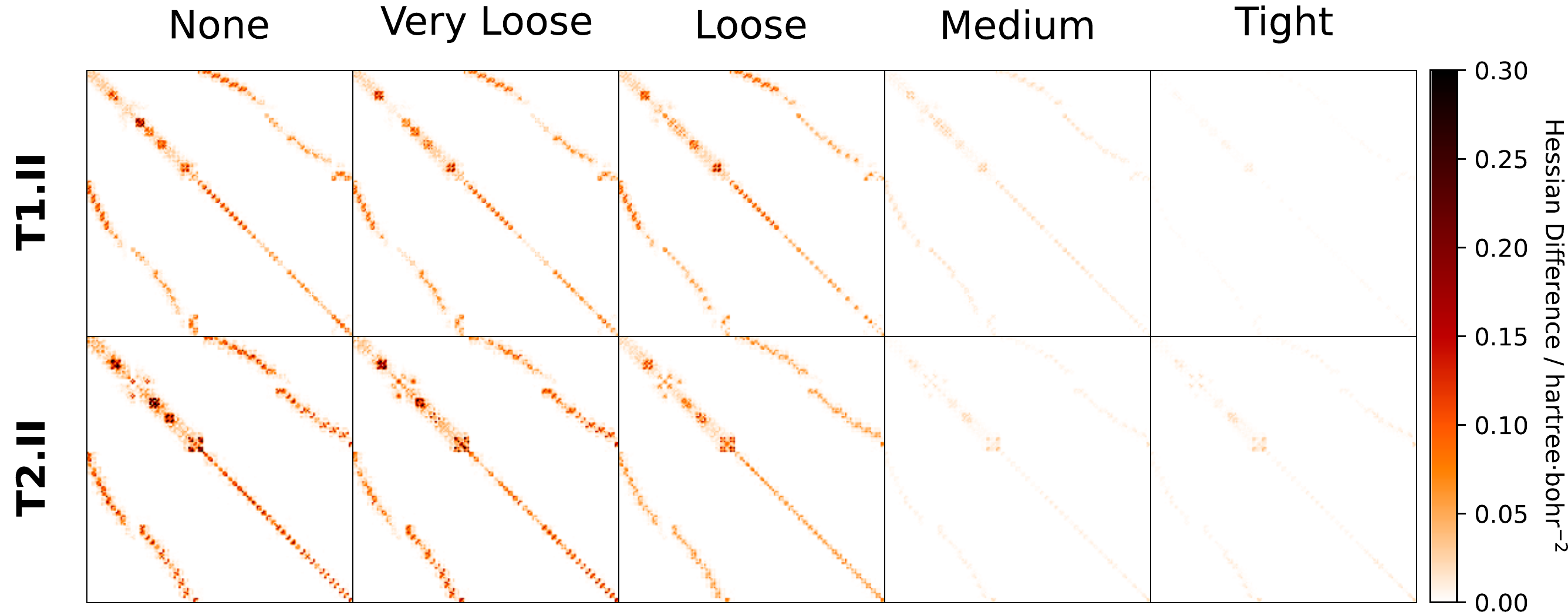}
    \caption{Top panel: from left to right, the panels correspond to the absolute deviations of the DFTB3 Hessian matrix elements of \textbf{T1.II.N}, \textbf{T1.II.VL}, \textbf{T1.II.L}, \textbf{T1.II.M}, and \textbf{T1.II.T} from the ones of \textbf{T1.II.VT}. Bottom panel: from left to right, the panels correspond to the absolute deviations of the DFTB3 Hessian matrix elements of \textbf{T2.II.N}, \textbf{T2.II.VL}, \textbf{T2.II.L}, \textbf{T2.II.M}, and \textbf{T2.II.T} from the ones of \textbf{T2.II.VT}. Each pixel in a panel corresponds to an element of the Hessian matrix, darker colors correspond to larger deviations. The calculated matrices are sparse, therefore most of the matrix elements have a negligible absolute deviation.}
    \label{fig:approxopt_hessian}
\end{figure}

In order to identify the sets of optimization criteria that still yield an acceptable accuracy, we compared the elements of the Hessian matrices obtained by the differently optimized molecular structures with the ones obtained from a structure optimized with the ``Very Tight'' convergence criteria. The deviation in the Hessian matrix elements in Cartesian coordinates of \textbf{T1.II} and \textbf{T2.II} are depicted in Fig.~\ref{fig:approxopt_hessian}. Even though the nature of the rearrangement is different, both molecular systems show a similar trend with increasing tightness of the convergence criteria: the deviation in the Hessian elements are strong in \textbf{T1.II.N} and \textbf{T2.II.N} and similarly present in \textbf{T1.II.VL} and \textbf{T2.II.VL}. In the Hessian matrices calculated from \textbf{T1.II.L} and \textbf{T2.II.L} the deviations diminish, especially noticeable in \textbf{T2.II.L}, and are negligible in the Hessian calculated from \textbf{T1.II.M} and \textbf{T2.II.M}. 

\begin{table}[hbpt]
    \centering
        \caption{RMSD of the calculated vibrational frequencies obtained for the structure optimized with different convergence criteria and compared to the ones obtained after optimization with the ``Very Tight'' convergence profile. The IR spectral region is divided into a low frequency region ($< 800$\,cm$^{-1}$), a middle frequency region (between 800\,cm$^{-1}$ and 2000\,cm$^{-1}$), and a high  frequency region ($> 2000$\,cm$^{-1}$). The vibrational frequencies are calculated for the second minimum along the trajectories \textbf{T1} and \textbf{T2} with the DFTB3 Hamiltonian.}
        \vspace{0.1cm}
%    \small
    \tabcolsep=0.07cm
    \begin{tabular}{l|lccc}
    \hline\hline
            &  Optimization  & Low $\tilde{\nu}_p$ & Middle $\tilde{\nu}_p$ & High $\tilde{\nu}_p$ \\
         Structure & profile & RMSD [cm$^{-1}$]&  RMSD [cm$^{-1}$] & RMSD [cm$^{-1}$] \\
         \hline
         \textbf{T1.II} & None       & 47.7 & 20.5 & 22.3\\
                     & Very Loose & 16.1 & 6.0  & 2.1\\
                     & Loose      & 9.8  & 4.6  & 3.1\\
                     & Medium     & 3.14  & 1.5 & 1.6\\
                     & Tight      & 0.3  & 0.2  & 0.2\\
                     \hline
        \textbf{T2.II}  & None       & 45.9 & 23.0 & 47.0\\
                     & Very Loose & 12.6 & 5.2  & 2.6\\
                     & Loose      & 7.5  & 3.4  & 1.5\\
                     & Medium     & 1.4  & 0.5  & 0.5\\
                     & Tight      & 1.1  & 0.4  & 0.5\\
        \hline\hline
    \end{tabular}
    \label{tab:freq_rmsd_approxopt}
\end{table}

Comparing the elements of the Hessian matrix allows us to conservatively evaluate the error introduced because the Hessian matrix is calculated from structures that have been differently optimized and therefore are in slightly different conformations. The comparison of the Hessian elements expressed in Cartesian coordinates leads to spurious differences due to local rotations in the molecular structure. In this case, even if the effect on the normal mode vibrational frequency is negligible, the effect on the blocks of the Hessian matrix corresponding to the molecular fragments involved in the local rotation are sizeable. We therefore inspected the difference in the vibrational frequencies, summarized in Table~\ref{tab:freq_rmsd_approxopt}, which turned out to be robust with respect to the above-mentioned spurious effects in the Hessian matrix elements.

We assigned the normal modes of the different calculations according to their energetic order, because carrying out a linear sum assignment with the absolute value of the Duschinsky matrix suffers from the fact that the overlap of modes expressed in Cartesian coordinates is no appropriate similarity metric for different molecular structures.
Comparing the vibrational frequencies obtained from structures optimized with different convergence criteria shows that there are distinct differences between how the normal modes in the low and middle spectral regions ($< 2000$\,cm$^{-1}$) behave compared to the normal modes in the higher frequency region ($> 2000$\,cm$^{-1}$). While for normal modes lying in the lower spectral region a RMSD in the vibrational frequencies smaller than 5\,cm$^{-1}$, suitable for a qualitatively reliable spectrum, is reached only from a structure optimization with the ``Medium'' optimization profile, for the stiff modes this accuracy is reached already with the structure optimized with the ``Very Loose'' convergence criteria. Furthermore, the error in the frequencies in the higher IR spectral region decreases faster than the respective error in the Hessian matrix elements. Molecular fragments involved in localized, stiff modes, such as -CH or -OH stretching, usually found at the high-frequency end of the IR spectrum, are easier to optimize than low-frequency normal modes which are often delocalized across the whole molecule, or internal rotations of fragments of comparably great size.

This analysis highlights three important facts of partially optimizing a molecular structure prior to the calculation of an IR spectrum. First, the computational time can be significantly lowered by adopting looser convergence thresholds. Second, if necessary, the error introduced by such optimizations can be efficiently reduced by tightening the structure optimization convergence criteria. Third, for the spectral region including the diagnostic IR spectral bands ($> 2000$\,cm$^{-1}$), speedups by up to one order of magnitude are possible while keeping the RMSD of the frequencies below 5\,cm$^{-1}$.

\subsubsection{Partial Hessian Approach}

\begin{table}[hbpt]
    \centering
        \caption{Summary of the partial Hessian approach for the \textbf{T1} trajectory. The time needed to calculate the Hessian matrix as well as the number of nuclei for which the second derivative of the energy with respect to the nuclear Cartesian coordinates need to be evaluated is shown for \textbf{T1.II} for different thresholds $\epsilon_{\rm RMSD}$. The time is indicated as mean $\pm$ standard deviation in milliseconds. The IR spectral region is divided in a low frequency region ($< 800$\,cm$^{-1}$), a middle frequency region (between 800\,cm$^{-1}$ and 2000\,cm$^{-1}$), and a high  frequency region ($> 2000$\,cm$^{-1}$). The RMSD between the vibrational frequencies with every $\epsilon_{\rm RMSD}$ and the one with $\epsilon_{\rm RMSD} = 0$\,bohr (equivalent to no partial Hessian approach) is given for the three spectral regions. }
        \vspace{0.3cm}
%    \small
    \tabcolsep=0.2cm
    \begin{tabular}{lccccc}
\hline
\hline
            $\epsilon_{\rm RMSD}$  & Time & Nuclei to & Low $\tilde{\nu}_p$ & Middle $\tilde{\nu}_p$ & High $\tilde{\nu}_p$ \\
         $[$bohr$]$ & [ms] & Evaluate & RMSD [cm$^{-1}$]&  RMSD [cm$^{-1}$] & RMSD [cm$^{-1}$] \\
               \hline 
                0.0        & $2341 \pm 5$ & 60 & -   &  -   & -    \\
                0.05       & $1408 \pm 4$ & 32 & 108 & 10.6 & 0.77 \\
                0.1        & $753 \pm 1$  & 16 & 120 & 11.6 & 1.47 \\
                0.2        & $764 \pm 9$  & 13 & 86  & 12.1 & 1.42 \\
                0.3        & $511 \pm 2$  & 9  & 18  & 7.0  & 1.17 \\
                0.5        & $498 \pm 5$  & 7  & 165 & 145  & 248  \\
\hline
\hline
    \end{tabular}
    \label{tab:freq_rmsd_partial}
\end{table}

\begin{figure}[hbpt]
    \centering
    \includegraphics[width=\textwidth]{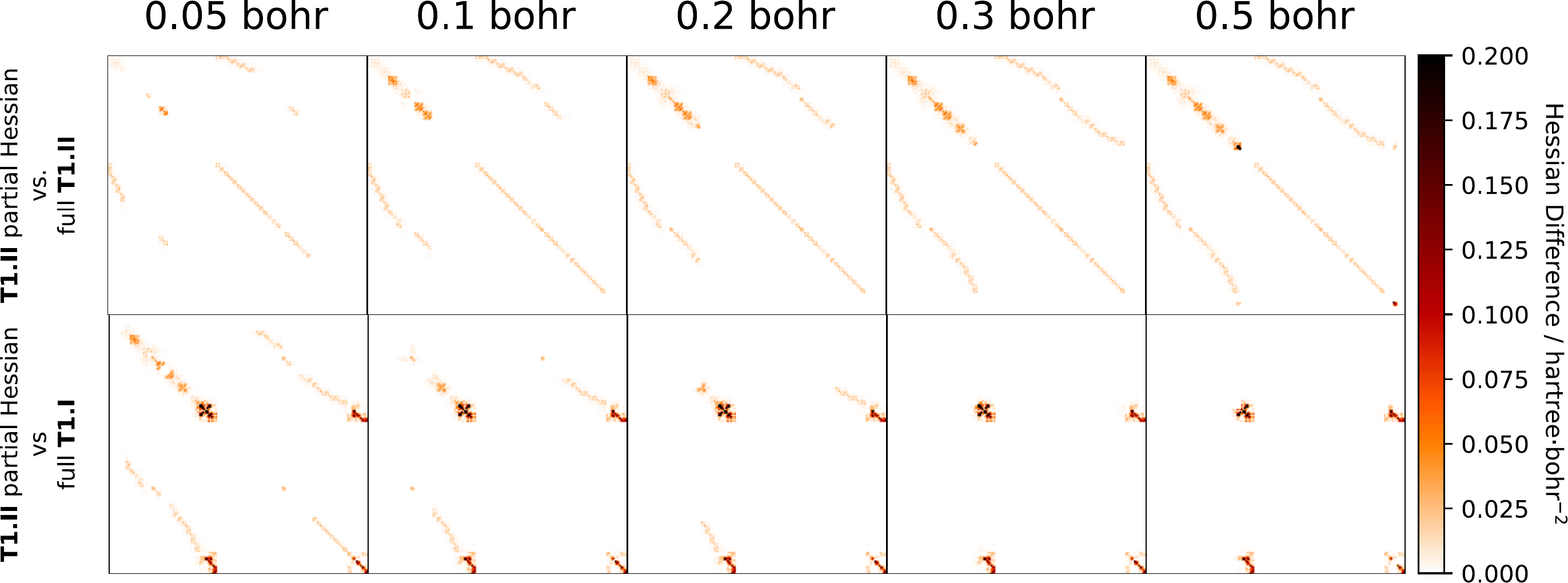}
    \caption{Comparison of the elements of the Hessian matrices with the partial Hessian approach for the second minimum along trajectory \textbf{T1}. From left to right: increasingly loose thresholds for the detection of structural fragments to recalculate, $\epsilon_{\rm RMSD}$ (0.05\,bohr, 0.1\,bohr, 0.2\,bohr, 0.3\,bohr, 0.5\,bohr). Top row: deviation of the elements of the Hessian matrix calculated in the second minimum along the trajectory compared with the Hessian matrix for the same structure but without the partial Hessian approximation. Bottom row: deviation of the elements of the Hessian matrix calculated in the second minimum along the trajectory compared with the Hessian matrix calculated in the first minimum. A difference is present only for the elements of the Hessian matrix corresponding to molecular fragments that have a RMSD determined by quaternion fitting larger than $\epsilon_{\rm RMSD}$.}
    \label{fig:t1_partial_hessian_compared}
\end{figure}

We assessed the partial Hessian approach at the example of the trajectories $\textbf{T1}$ and $\textbf{T2}$. The first minimum in both trajectories is calculated without any of the approximations introduced in this work. In the second minimum, only the elements of the Hessian matrix corresponding to fragments in the molecular structure that are not similar enough to the previous structure are evaluated. The rest of the Hessian matrix is copied from the one calculated in the previous minimum. A single parameter, $\epsilon_{\rm RMSD}$, controls the similarity threshold between the nuclear coordinates, and represents the maximum RMSD of the Cartesian coordinates of each nucleus after an iterative alignment. A least-square quaternion fitting of the molecular structures is not an ideal option for the identification of invariant structural fragments. Ideally, a local alignment algorithm would ignore the fragments of the molecule that were distorted and optimally fit the target molecule to the parts of the molecular structures that are not distorted between two neighboring local minima. In such a way the number of Hessian matrix elements that need to be reevaluated is kept at a minimum. 

The partial Hessian approach consists of 2 ingredients. First, the iterative alignment algorithm identifies the molecular fragments for which the chemical environment has significantly changed from the previous minimum, \textit{i.e.}, for which the corresponding Hessian matrix elements need to be recalculated. Second, the blocks of the Hessian matrix corresponding to the identified fragments are evaluated as numerical differences of analytical gradients. This is a trivially parallel task and can be implemented by adapting a full semi-numerical Hessian evaluation algorithm. The iterative alignment algorithm ensures that the local distortions from one minimum to the other one do not affect the parts of the molecule that were not affected by the local distortion. Common structural fragments may be difficult to identify in the case of local minima on a trajectory connected by a global distortion, and the whole Hessian matrix may need to be calculated. 

The partial Hessian approach is particularly advantageous for local minima connected to previous ones by localized structural distortions. This is corroborated by the data shown in Fig.~\ref{fig:t1_partial_hessian_compared} for trajectory \textbf{T1}, where the enol undergoing a keto--enol tautomerism is at the end of an aliphatic chain. Most of the molecular structure is largely unaffected by this rearrangement, and only a small fraction of the Hessian elements needs to be updated (the molecular fragments that need to be evaluated for each $\epsilon_{\rm RMSD}$ are indicated in the supplementary information). In the bottom row of Fig.~\ref{fig:t1_partial_hessian_compared}, the Hessian matrix calculated at a minimum along the trajectory \textbf{T1} is compared with the one calculated in the previous minimum. The molecular fragments responsible for the most intense deviations in the Hessian matrix elements with respect to the one of the previous minimum are readily identified and recalculated (Fig.~\ref{fig:t1_partial_hessian_compared}, bottom row), even at comparably high $\epsilon_{\rm RMSD}$. In this favorable example, the time needed to update the Hessian matrix is reduced from 2341\,ms in the full calculation to 511\,ms in the calculation with $\epsilon_{\rm RMSD} = 0.3$\,bohr. Higher thresholds lead to a severe misrepresentation of the normal modes and normal mode frequencies, as summarized in Table~\ref{tab:freq_rmsd_partial}. Normal modes were assigned through a linear sum assignment with the absolute value of the Duschinsky matrix as score matrix. The RMSD in the frequencies in the low and middle IR spectral regions is not minimal at the smallest $\epsilon_{\rm RMSD}$, but rather at a threshold of 0.3\,bohr. This is due to the fact that Hessian matrix elements important for delocalized normal modes are calculated at different minima. A higher fidelity is achieved by taking all the Hessian matrix elements at the same structure. Such delocalized normal modes are prominently present in the low frequency range of the spectra, and this explains why this effect is especially present in frequencies lower than 800\,cm$^{-1}$.

In \textbf{T2.II}, the rotation around a central dihedral after the tautomerization during the structure optimization causes a global structural rearrangement, and the alignment does not recognize any fragment which is similar enough to a fragment in the previous minimum (the alignment of the optimized structure is depicted in the supplementary information). In this case, the full Hessian matrix is calculated. Hence, if the character of the path connecting two local minima on a PES is that of a global structural rearrangement, the reliability of the Hessian matrix calculated is not affected, rather, the performance of the approach is. Two avenues could be explored to overcome the current limitation in the iterative alignment algorithm. First, the optimized structure corresponding to multiple local minima can be saved, and the structure at hand can be aligned to multiple previous structures to find the one with a better match. During an exploration, the probability to find similar structures grows with the number of structures that have been already explored. Second, the iterative alignment algorithm could be improved by first partitioning the molecular structure into local fragments that are independently aligned\cite{Brunken2020}.

A small difference in the time needed to evaluate the Hessian matrix with $\epsilon_{\rm RMSD} = 0.0$\,bohr and the time given in Table~\ref{tab:time_normal} is possible even though they result in the same number of matrix elements to calculate, as the algorithm for the evaluation of a partial Hessian matrix is slightly different to the one for the evaluation of the full Hessian matrix.

\subsection{UV/Vis Spectroscopy}

\begin{figure}[hbpt]
    \centering
    \includegraphics[width=\textwidth]{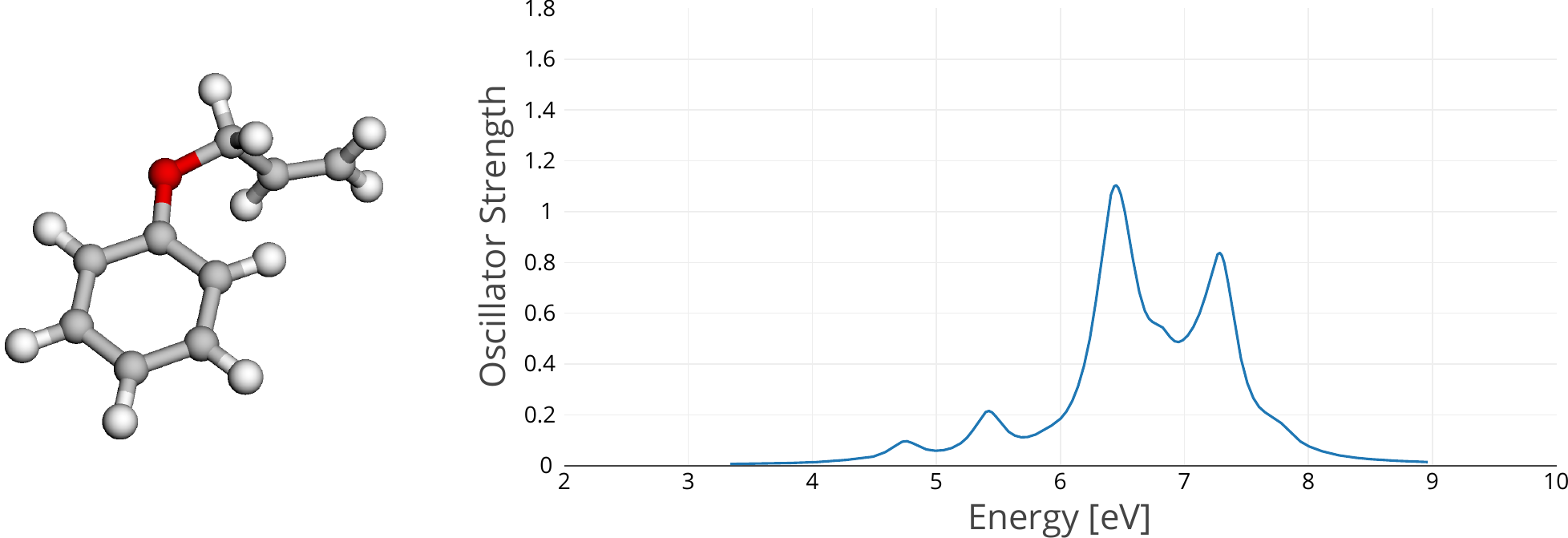}
    \caption{The interactive spectroscopy approach for the structures on the trajectory \textbf{MD}. On the left panel, a molecular structure is displayed. On the right panel, the corresponding UV/Vis spectrum is shown. In the HTML version of this work, this figure is available in an interactive format. One can indicate the desired structure along the trajectory in writing its index in the ``Index'' box or by moving the slider. Both, spectra and structures, are available as javascript arrays for interactive use in the online version of this paper.}
    \label{fig:md_interactive_static}
\end{figure}

\begin{figure}[hbpt]
    \centering
    \includegraphics[width=\textwidth]{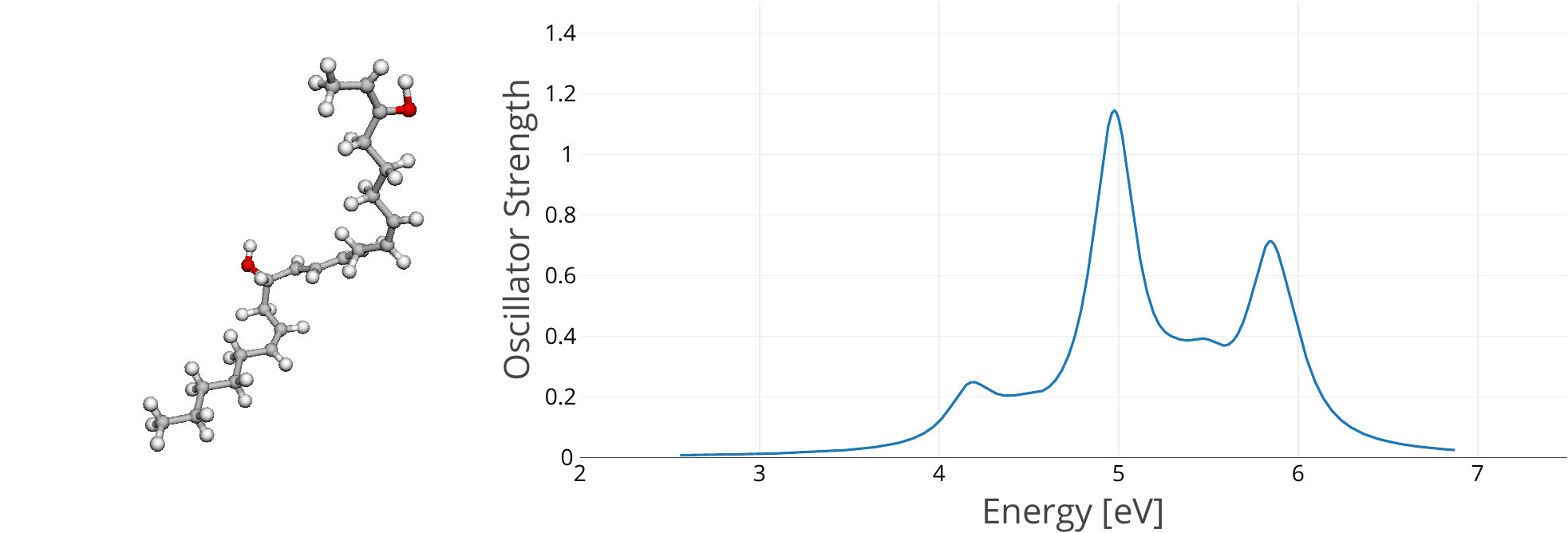}
    \caption{The interactive spectroscopy approach for the structures on the trajectory \textbf{T1}. On the left panel, a molecular structure is displayed. On the right panel, the corresponding UV/Vis spectrum is shown. In the HTML version of this work, this figure is available in an interactive format. One can indicate the desired structure along the trajectory in writing its index in the ``Index'' box or by moving the slider. Both, spectra and structures, are available as javascript arrays for interactive use in the online version of this paper.}
    \label{fig:t1_interactive_static}
\end{figure}

We calculated the UV/Vis spectra for the trajectories \textbf{MD} and \textbf{T1}, \textit{cf.}, Figs.~\ref{fig:md_interactive_static} and~\ref{fig:t1_interactive_static} (in the interactive HTML version of this work, the spectrum of every structure along both trajectories can be inspected). While the spectrum for the \textbf{MD} system behaves rather eratic along the trajectory due to the comparatively large structural changes, the power of real-time UV/Vis spectroscopy for diagnostic purposes is very well illustrated by the \textbf{T1} system. Here, a small peak at about 6\,eV is present for the alcohol. This peak sharply increases in intensity around the transition state (see Fig.~\ref{fig:t1_interactive_static}), only to vanish completely for the ketone at the end of the trajectory.

\begin{figure}[hbpt]
    \centering
    \includegraphics[width=\textwidth]{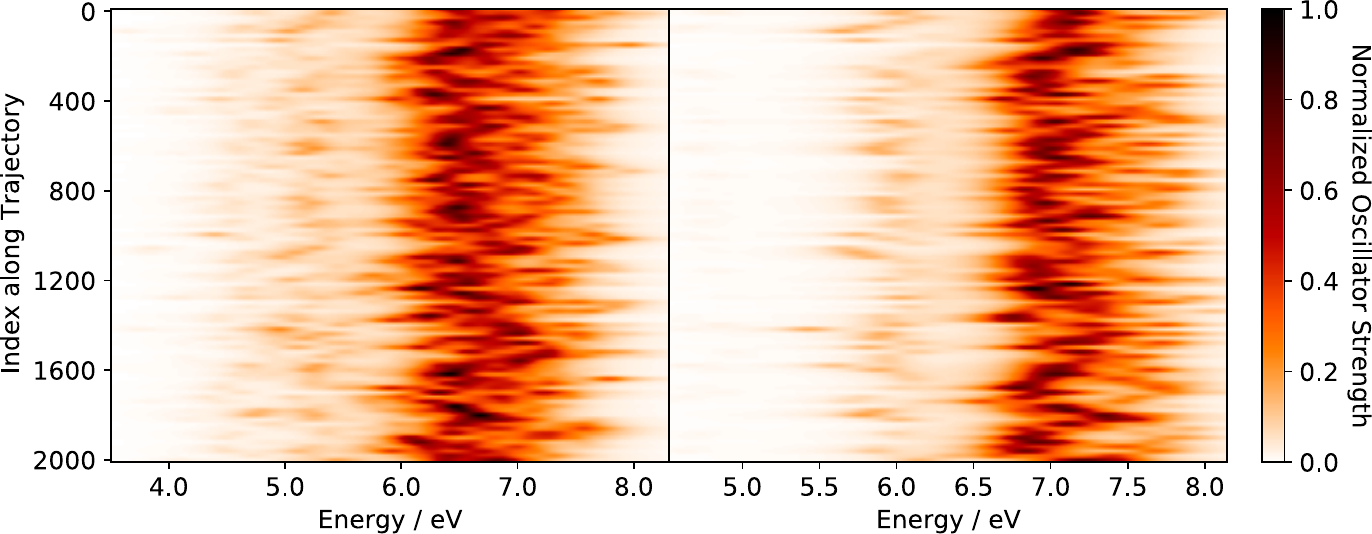}
    \caption{Comparison of the UV/Vis spectra of TD-DFTB3 and linear-response SCS-CC2 for a subset of structures along the \textbf{MD} trajectory. Left panel: the first 30 electronic excited states calculated with the TD-DFTB3 method are convoluted with a Lorentzian function with full-width at half-maximum of 0.3\,eV. Right panel: the first 10 electronic excited states calculated with the linear-response SCS-CC2 method. Darker colors correspond to more intense spectral bands, and the color is given by the oscillator strength of the electronic transition normalized to the one of the most intense electronic transition for each method. Every horizontal projection is the UV/Vis spectrum of a single structure. One every twenty structures along the trajectory was sampled for calculation. A total of 101 structures was calculated.}
    \label{fig:dftb3_vs_cc2}
\end{figure}

\begin{figure}[hbpt]
    \centering
    \includegraphics[width=0.65\textwidth]{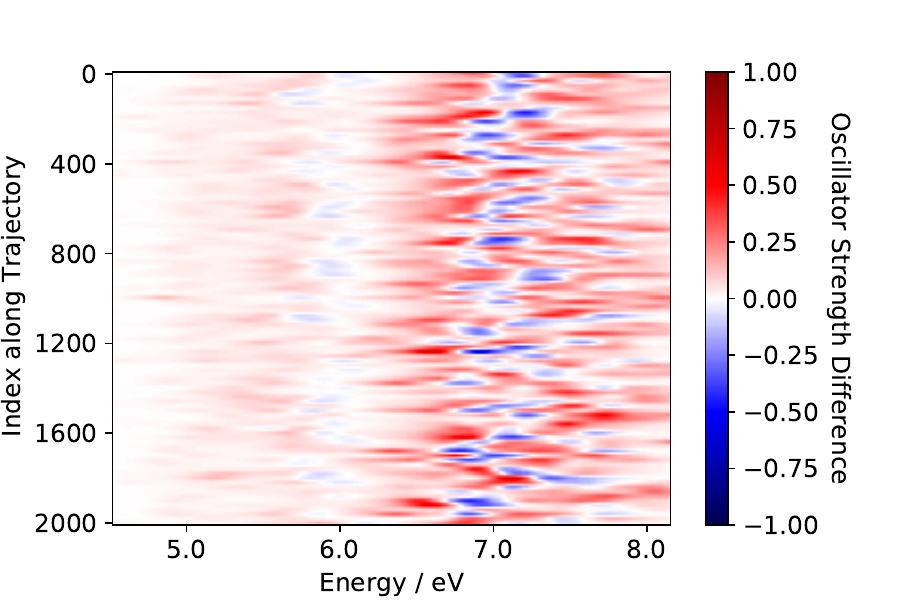}
    \caption{Difference between the UV/Vis spectra of TD-DFTB3 and linear-response SCS-CC2 for a subset of the \textbf{MD} trajectory, with a constant blue-shift of 0.46\,eV applied to the TD-DFTB spectrum. The spectrum of TD-DFTB is calculated by convolution of the first 30 excited states, the SCS-CC2 one by convolution of the first 10 excited states with a Lorentzian function with full-width at half-maximum of 0.3\,eV. Blue parts of the spectrum indicate that the normalized oscillator strength of the SCS-CC2 spectrum is larger than the one of TD-DFTB, and vice-versa for red colors. Darker colors correspond to larger differences. Every horizontal projection is the difference UV/Vis spectrum of a single structure. One every twenty structures along the trajectory was sampled for calculation. A total of 101 structures were calculated.}
    \label{fig:dftb3_vs_cc2_diff}
\end{figure}

An appropriate method for efficient electronic excited-state calculations must yield qualitatively comparable results to more accurate methods at a fraction of the cost. In Fig.~\ref{fig:dftb3_vs_cc2}, we show that through the calculation of enough states, the linear-response SCS-CC2 spectrum is recovered qualitatively by the TD-DFTB method, albeit being red-shifted by 0.46\,eV. The difference spectrum in Fig.~\ref{fig:dftb3_vs_cc2_diff} also shows that the difference between the two spectra after blue-shifting the TD-DFTB spectra by 0.46\,eV is acceptable, in light of the fact that transition properties as oscillator strengths are particularly sensitive, with a mean of the maximum absolute deviation of the normalized intensity in each spectrum of 0.35. The results of a PBE/def2-TZVP/TD-DFT calculation are similar to the ones obtained with TD-DFTB, and the same calculation with the PBE0 hybrid exchange--correlation functional cures in part the ghost-state problem (data provided in the supplementary information).

The adequacy of the initial guess is of importance for the convergence properties of the subspace solver. Therefore, we attempted to devise two approaches to provide starting vectors to the iterative diagonalizer that are closer to the solution of the excited-state problem. First, the initial guess was provided by the solution of the excited-state problem of the previous structure along the trajectory under study. Second, a linear combination of previous excited-state solutions along the lines of the DIIS approach we introduced\cite{Muehlbach2016} for the acceleration of the self-consistence-field convergence in ground-state calculations was attempted. Both strategies showed a limited acceleration of the calculation of the first 30 excited states for each structure along the trajectory \textbf{T1} with an initial guess provided by one of the two previous strategies compared to the standard guess of Eq.~(\ref{eq:init_guess}) (see supplementary information). However, this effect was not observed anymore if the same initial subspace was complemented by 90 standard initial vectors for a total of 120 trial vectors.

The comparison of the efficiency of our implementation against the one of the DFTB+ software package for the calculation of the first 30 excited states with an initial guess space of 30 vectors of the first structure of the \textbf{T1} trajectory revealed that the average total wall time required by the DFTB+ program for the excited-state calculation was 1.0\,s, whereas for the same calculation the average wall time required by \textsc{Sparrow} was 246\,ms (wall time obtained as an average over 3 calculations).

\subsubsection{Pruning the Excited-State Basis}

\begin{figure}
    \centering
    \includegraphics[width=\textwidth]{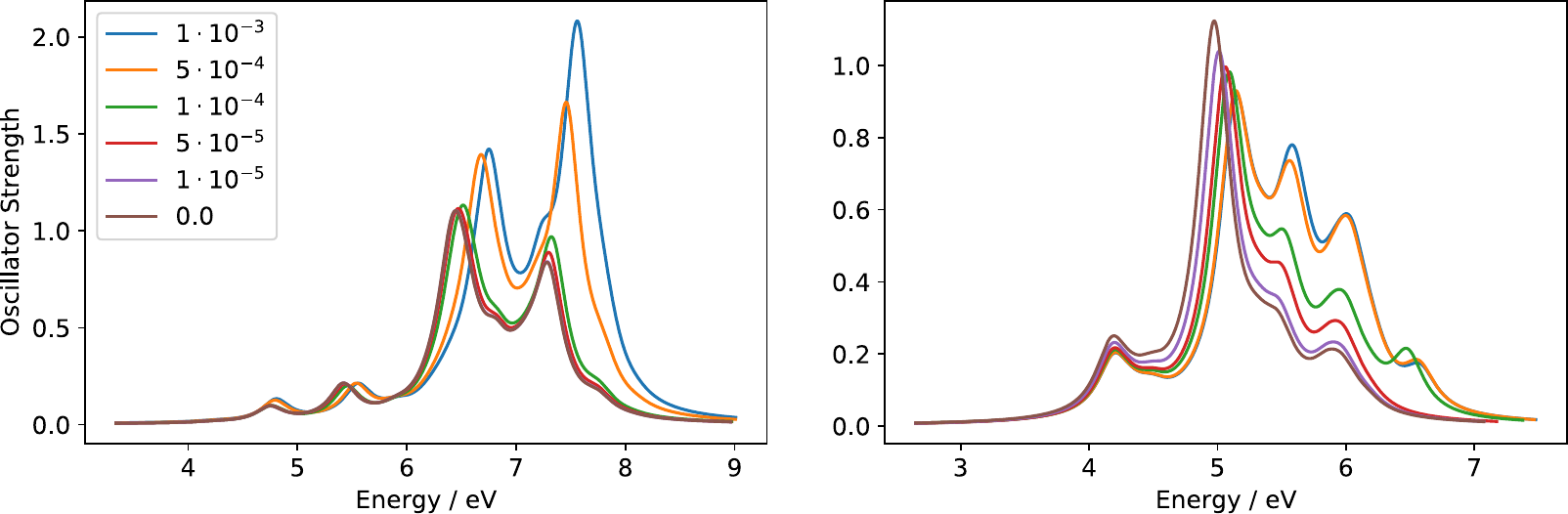}
    \caption{Comparison of the UV/Vis spectra calculated with TD-DFTB3 and different $\epsilon_{\rm PT2}$ for the first structures of the trajectories \textbf{MD} and \textbf{T1}. For both systems, the first 30 electronic excited states of the first structure along the respective trajectory are calculated with the TD-DFTB3 method and are convoluted with a Lorentzian function with full width at half maximum of 0.3\,eV. Left panel: resulting UV/Vis spectra for the first structure along the \textbf{MD} trajectory. Right panel: resulting UV/Vis spectra for the first structure along the \textbf{T1} trajectory. The inclusion thresholds described in the theory section for each spectrum are indicated in the legend in units of hartree$^2$. If this threshold was zero, the whole excited-state basis had been included.}
    \label{fig:md_t1_pruned}
\end{figure}

Especially for systems with many possible electronic transitions, the improved iterative diagonalizer alone may be insufficient to provide the required acceleration. Therefore, we assessed the suitability of approximate solutions of the excited-state problem through the limitation of the size of the excited-state basis. This approach exploits the diagonally-dominant structure of the matrix $\boldsymbol{\Delta}^\frac{1}{2} (\boldsymbol{A} + \boldsymbol{B}) \boldsymbol{\Delta}^\frac{1}{2}$ in Eq.~(\ref{eq:tddftb}). Including only as many determinants with the lowest diagonal component as solutions required in the excited-state problem neglects the coupling between these determinants and the rest of the excited-state space. This issue could be tamed by additionally including all determinants that couple with the first determinants according to the criterion derived by perturbation theory described in Eq.~(\ref{eq:pert_crit}) more than a threshold $\epsilon_{\rm PT2}$. In Fig.~\ref{fig:md_t1_pruned}, we compare UV/Vis spectra obtained with several $\epsilon_{\rm PT2}$ for the first structure of the trajectories \textbf{MD} and \textbf{T1}. As expected, the smaller the inclusion threshold $\epsilon_{\rm PT2}$ becomes, the better the full spectrum is described. For limits below $10^{-4}$\,hartree, almost no difference to the exact spectrum can be made out. Moreover, the first peaks of the two spectra are less dependent on $\epsilon_{\rm PT2}$ than the ones at the higher end of the spectra. The roots responsible for the first peaks are allowed to couple with the other determinants with a diagonal element lower than the maximally required energy span of the spectrum independently from $\epsilon_{\rm PT2}$. Hence, these states are often well described already without any additional basis function, provided a sufficient number of excited states is to be determined.

In Table~\ref{tab:time_t1_md_pruning}, we provide the number of basis functions and the timings required to calculate the first 30 excited states with an initial subspace dimension of 30 of the first structure in the trajectories \textbf{MD} and \textbf{T1} with several $\epsilon_{\rm PT2}$. While the accuracy is not very dependent on the size, the computational gain of pruning the excited-state space is. Calculating the first 30 excited states of a structure in the \textbf{MD} trajectory with sufficient accuracy ($\epsilon_{\rm PT2} = 5\cdot 10^{-5}$\,hartree$^2$) is a modest two times faster than without pruning. The speedup will grow if the calculation is carried out for a larger system. In fact, the same spectrum for the first structure of the \textbf{T1} trajectory is calculated already 3.4 times faster than the respective calculation in the full space. 

The reasons of the speedup are twofold: first, all linear algebra operations, such as the generation of the sigma vectors, are now carried out in a smaller space. Second, the number of iterations of the Davidson algorithm is smaller. The reduction of the dimension of the excited-state basis can potentially allow for the efficient non-iterative diagonalization of the matrix in the eigenvalue problem in Eq.~(\ref{eq:tddftb}). 

\begin{table}[hbpt]
    \centering
        \caption{Dimension of the excited-state basis, mean computational time and speedup for the calculation of the first 30 transitions of the UV/Vis spectra corresponding to the first structures along the trajectories \textbf{MD} and \textbf{T1} with several $\epsilon_{\rm PT2}$ for the pruning of the excited-state basis. Timings are given as the mean of the time required to calculate an UV/Vis spectrum over 3 calculations. The speedup is relative to the time without pruning in the first column. Significant digits are given by the standard deviation: if it is larger than 2.5 multiplied by the appropriate power of ten, then it is rounded to the first digit, otherwise to the second one.}
        \vspace{0.1cm}
    \begin{tabular}{l|cccccc}
\hline
\hline
               & \multicolumn{6}{c}{$\epsilon_{\rm PT2}$ [hartree$^2$]}  \\
              System        & 0 & $1\cdot 10^{-5}$ & $5\cdot 10^{-5}$ & $1\cdot 10^{-4}$ &      $5\cdot 10^{-4}$ & $1\cdot 10^{-3}$\\
               \hline
         \textbf{MD} &&&&&&\\
         \hline
          Dimension & 634 & 471 & 288 & 182 & 53 & 38 \\
          Time [ms] &$39.3 \pm 0.9$ & $40 \pm 3$ & $29.3 \pm 0.5$ & $29.3\pm 1.2$ & $12 \pm 3$ & $9.7 \pm 0.4$ \\
          Speedup &1x&1.3x&2.0x& 2.5x &9.0x&8.8x\\
         \hline
         \textbf{T1} &&&&&&\\
         \hline
         Dimension & 4352 & 848 & 252 & 151 & 45 & 35\\
         Time [ms] & $259.0 \pm 0.8$ & $116 \pm 6$ & $84.0 \pm 0.8$ & $52 \pm 6$ & $30.0 \pm 0.0$ & $29.0 \pm 0.0$\\
         Speedup & 1x & 2.2x & 3.1x & 5.0x & 8.6x & 8.9x\\
\hline
\hline
    \end{tabular}
    \label{tab:time_t1_md_pruning}
\end{table}

\clearpage

\section{Conclusions}

Computational spectroscopy in high-throughput and interactive quantum chemistry settings is challenging due to its high computational cost. Even with suitably parametrized models, such as semi-empirical Hamiltonians, obtaining spectroscopic information with sufficient accuracy at a high rate is a formidable task that requires the development of tailored approaches for the reduction of computational hurdles. 

The approaches discussed in this work allow for the efficient calculation of spectroscopic signals in high-throughput and interactive quantum chemistry. While some of these methods are specific for the calculations of closely related structures, others are of more general applicability. 

Vibrational spectroscopy in the harmonic approximation presents two computational bottlenecks: structure optimization and Hessian-matrix calculation. We pursued two options to accelerate these calculations. First, we assessed the viability of incomplete structure optimizations for the calculation of vibrational spectra. 
At an example, we characterized how different tightness of convergence thresholds for structure optimization affects the error in the spectroscopic peak positions and intensities and in the Hessian matrix elements. We identified a set of convergence criteria that were sufficient to reduce the computational time at a limited toll on accuracy. In particular, the diagnostic high-frequency spectral bands were well reproduced already with an approximate structure optimization due to the localized nature of the corresponding normal modes. 

Second, we introduced a partial Hessian approach to reduce the number of Hessian matrix elements to be calculated by leveraging the similarities between the structures corresponding to the local minima on the PES for which a spectrum is required. In order to do so, the structure corresponding to the local minimum for which a spectrum needs to be evaluated is compared with the one of the previous minimum. A local iterative alignment scheme, controlled by a single parameter, was designed to identify the invariant parts of the molecule. The elements of the Hessian matrix corresponding to parts of the molecular structure that have been successfully aligned and are therefore sufficiently similar are not recalculated but inherited from the previous structure. 

The application of these two approaches allowed for the acceleration of vibrational spectroscopy under control of the tolerable error.
The approximations introduced for the calculation of infrared spectra are particularly reliable for high-frequency, stiff normal modes. The localization of these vibrational modes also makes the two approaches more transferable to different molecular systems, as these modes are then less dependent on their chemical environment.

UV/Vis spectroscopy requires efficient methods for recovering sufficiently accurate vertical electronic transition energies and corresponding oscillator strengths. 
To tackle the high computational cost of the linear-response excited-state calculation, we implemented a non-orthogonal modification of the Davidson algorithm\cite{Parrish2016, Furche2016}. Furthermore, we devised a strategy to leverage this similarity by improving the initial guess of the iterative diagonalization, which we obtained as a linear combination of previous solutions of the excited-state problem with the DIIS algorithm\cite{Muehlbach2016}. However, the improved initial guess did not consistently decrease the time needed to reach a solution. A complementary approach is to solve the excited-state problem in a limited excited-state determinant space. By neglecting all determinants that are not coupling considerably with the solution subspace, the size of the excited-state problem could be massively reduced with limited accuracy losses that can be controlled by a single parameter.

Even though the approaches implemented in this work are primarily intended for the ultra-fast application with semi-empirical Hamiltonians, they are agnostic to the electronic structure model; i.e., they can also be applied to accelerate calculations with more accurate and computationally expensive methods. The application to semi-empirical models based on the neglect of diatomic differential overlap, such as MNDO, AM1, RM1, PM3 and PM6, yields, however, no reliable vibrational and electronic spectra (the latter calculated with the configuration interaction singles method). The extension of the approaches discussed in this work to modern semi-empirical models, such as the extended tight-binding method family (GFNn-xTB, n = 0, 1, 2) and orthogonalization-corrected methods (OMn, n = 1, 2, 3), is rather straightforward and will therefore be considered in future work.

\section*{Acknowledgments}

We gratefully acknowledge financial support by the Swiss National Science Foundation (Project No.~200021\_182400).
We thank Dr.~Alain Vaucher for discussions at the beginning of this work in 2018.

%\bibliography{RTSpec}
\providecommand{\latin}[1]{#1}
\makeatletter
\providecommand{\doi}
  {\begingroup\let\do\@makeother\dospecials
  \catcode`\{=1 \catcode`\}=2 \doi@aux}
\providecommand{\doi@aux}[1]{\endgroup\texttt{#1}}
\makeatother
\providecommand*\mcitethebibliography{\thebibliography}
\csname @ifundefined\endcsname{endmcitethebibliography}
  {\let\endmcitethebibliography\endthebibliography}{}

\end{document}